\documentclass[english,11pt]{article}
\usepackage[utf8]{inputenc}
\usepackage{graphicx}
\usepackage{color}
\usepackage{float}
\usepackage{slashed}
\usepackage{tocloft}
\usepackage{amsthm,amsmath,amssymb,mathrsfs}
\usepackage{setspace}
\allowdisplaybreaks
\usepackage{jheppub}
\usepackage{xcolor} 
\usepackage{babel}
\usepackage{booktabs}

\def\beq{\begin{equation}}
\def\eeq{\end{equation}}

\renewcommand{\Im}{{\mbox{\rm Im}}}
\usepackage{subcaption}
\usepackage[normalem]{ulem}

\newcommand{\vecp}{\mathbf{p}}
\newcommand{\vecr}{\mathbf{r}}

\newcommand{\sgn}{\text{sgn}}

\begin{document}
\title{Strongly Coupled Quantum Forces}
\author[a]{Yuval Grossman,}
\emailAdd{yg73@cornell.edu}
\author[a]{Chinhsan Sieng,}
\emailAdd{cs2284@cornell.edu}
\author[b]{Xun-Jie Xu,} 
\emailAdd{xuxj@ihep.ac.cn}
\author[a]{Bingrong Yu}
\emailAdd{bingrong.yu@cornell.edu}
\affiliation[a]{Department of Physics, LEPP, Cornell University, Ithaca, NY 14853, USA}
\affiliation[b]{Institute of High Energy Physics, Chinese Academy of Sciences, Beijing 100049, China}

\abstract{
Quantum forces are long-range interactions originating from vacuum fluctuations of mediator fields. Such forces inevitably arise between ordinary matter particles whenever they couple to light mediator species. Conventional computations of quantum forces rely on evaluating one-loop Feynman diagrams of the relevant scattering processes.
    In this work, we introduce a novel framework to compute quantum forces. Instead of relying on perturbative scattering amplitudes, we directly evaluate the quantum fluctuations of the mediator field by solving its quantized equation of motion with appropriate boundary conditions. This approach remains valid beyond the Born approximation and thus applies to regimes of strong coupling between the mediator and matter fields. In the weak-coupling limit, our results reproduce the known expressions from the Feynman diagram approach. In the strong-coupling regime, the result is modified by a factor that can suppress or enhance the effect. In contrast to classical forces, quantum forces intrinsically violate the superposition principle. Our approach may therefore offer a useful tool for probing non-perturbative effects in the infrared regime.
}

\maketitle

\section{Introduction}
\label{sec:intro}

The concept of force has a long and evolving history. Initially, it was a fundamental idea in Newtonian mechanics, describing interactions between masses. With the development of Maxwell equations and classical electrodynamics, the notion of force took on a new perspective: forces were understood as arising from fields generated by sources, with other particles responding to these fields.

Quantum mechanics further deepened our understanding of the concept of force. In particular, quantum field theory (QFT) provides a framework in which forces are mediated by gauge bosons. For example, the electromagnetic interaction is carried by photons. To recover classical forces from QFT, one typically computes Feynman diagrams, specifically, $t$-channel diagrams, which, when Fourier-transformed to real space, reproduce the familiar form of classical forces.

This approach naturally extends the notion of force into the quantum regime. A ``quantum force'' can emerge from $t$-channel diagrams containing loops rather than being restricted to tree-level interactions. In such cases, the classical field vanishes, yet a force arises purely from quantum effects. Perhaps the most famous example is the Casimir effect. In 1947, Casimir and Polder considered the electromagnetic interaction between two neutral atoms separated by a distance $r$ and derived the two-photon exchange potential~\cite{Casimir:1947kzi}
\begin{align}
V_{2\gamma}(r) = -\frac{23}{4\pi} \frac{\alpha_1\alpha_2}{r^7}\;,
\end{align}
where $\alpha_{i}$ is the polarizability of each atom.\footnote{ 
The more well-known form of the Casimir force between two perfectly conducting plates, with the potential energy per area ${\cal V}(r) = -\pi^2/(720\,r^3)$, was actually derived one year later by Casimir~\cite{Casimir:1948dh}.  The force was observed experimentally in 1997~\cite{Lamoreaux1997}.
}
This force is absent at the classical level, since an isolated atom carries no permanent dipole moment in the absence of external electromagnetic fields. Quantum mechanically, however, electromagnetic field fluctuations induce an effective interaction between neutral atoms. In the language of Feynman diagrams, the Casimir-Polder force corresponds to the one-loop process mediated by two virtual photons, which can be computed using perturbative  QFT~\cite{Feinberg:1989ps}.

Another well-known example of a quantum force is the neutrino force, generated by the exchange of two neutrinos~\cite{Feinberg:1968zz}. The effective potential of this force, assuming massless neutrinos, is 
\begin{align} V_{2\nu}(r) \sim \frac{G_F^2}{r^5}\,, \label{eq:nu-force}
\end{align}
where $G_F$ is the Fermi constant. This force, as an interesting prediction of the Standard Model (SM), has drawn considerable attention since it was  proposed~\cite{Hsu:1992tg,Horowitz:1993kw,Grifols:1996fk,Fischbach:1996qf,Smirnov:1996vj,Abada:1996nx,Kiers:1997ty,Ferrer:1998ju,Ferrer:1998rw,Ferrer:1999ad, Lusignoli:2010gw,
 Segarra:2015mqp, Stadnik:2017yge, Asaka:2018qfg, Ghosh:2019dmi, LeThien:2019lxh,  Segarra:2020rah,Costantino:2020bei,Orlofsky:2021mmy, Xu:2021daf,Coy:2022cpt,Ghosh:2022nzo,Blas:2022ovz,VanTilburg:2024xib,Ghosh:2024qai,Ghosh:2024ctv}.
As the third example, the two-graviton-exchange potential, which constitutes the leading quantum correction to the classical Newtonian potential in the infrared (IR) regime, has been computed both for point-like particles~\cite{Donoghue:1993eb,Donoghue:1994dn,Donoghue:1995cz} and for extended objects such as cosmological strings~\cite{Wessling:2001jb}.
Going beyond the SM, quantum forces have been calculated for a variety of mediators, such as axions~\cite{Ferrer:1998ue,Bauer:2023czj,Grossman:2025cov}  and other exotic light particles in the dark sector \cite{Fichet:2017bng,Brax:2017xho}. 

Traditionally,  calculations of quantum forces have been performed using leading-order Feynman diagrams with the Born (perturbative) approximation, similar to how the Coulomb potential is derived in QFT~\cite{Peskin:1995ev}, except that 
the former involves loop diagrams and the latter uses tree-level diagrams. 
However, such calculations lack an intuitive picture of how quantum forces are generated. Moreover, the Born approximation is not always valid, as it involves non-relativistic scattering in quantum mechanics and assumes weak coupling between the mediator and matter fields. It is important to recall that the Coulomb potential can be derived in classical electrodynamics without involving any Feynman diagrams, nor the Born approximation. Hence, it would be desirable if quantum forces can be derived beyond the diagrammatic approach, which is the goal of this paper.

Our method offers an intuitive perspective by treating quantum forces as a type of background effect. In the classical picture, forces such as the Coulomb force arise from a background field (e.g., the electric field) that influences the motion of a particle placed within it. For quantum forces, the classical field itself vanishes, but vacuum fluctuations play the role of effective force carriers. In this sense, the vacuum behaves analogously to a classical background, and our method is close in spirit to the classical field picture. 

Motivated by this analogy, we extend background-field techniques to the vacuum case by incorporating expectation values of quantum fields. Specifically, we extend the notion of a background field into the quantum regime by taking expectation values that classically vanish. This method not only reproduces the known results obtained from the diagrammatic approach in the weak-coupling limit but also allows us to analytically access regimes where the Born approximation breaks down. The key advantage of our method over previous approaches is precisely this ability to treat situations in which the mediator field couples strongly to matter.

Below, we clarify the notion of ``strong coupling'' used in this work. We emphasize that we only focus on the long-range (IR) behavior of the force throughout this paper. Our central idea is to probe the strongly coupled regime in the IR by exploiting the analogy between a large microscopic coupling and a high-density environment.

In electromagnetism, extending to large charges is straightforward. The potential between any two charges is given by Coulomb’s law, and in the IR, that is at large distances, only the monopole term matters. Thus, it does not matter whether the source consists of many small charges or a single large charge. Calculations in electromagnetism are simple: for many particles, one can just add their potentials. This is captured by the superposition principle: potentials from multiple sources add linearly, a consequence of the photon coupling linearly to the electron.

For general quantum forces, this linearity fails. Yet, in the IR, we still do not care about the details of the source. Thus, here we assume perturbativity at the fundamental level and consider many particles confined within a region of size $R$. Far from the source, that is at $r \gg R$, the detailed structure becomes irrelevant, allowing exploration of strong-coupling effects by increasing the density. 
Thus, in the following, we define ``strong coupling'' in two related ways: (1) as a high-density system of weakly coupled particles, or (2) as a point-like particle with a large effective coupling. The strongly coupled regime accessible through our method can then be mapped onto the non-perturbative coupling of the underlying field theory in the IR limit (see Sec.~\ref{subsec:strongcoupling-nonperturbative}).

In this paper, we focus on the simplest nontrivial examples of quantum forces, in which the mediator couples quadratically to matter. In such theories, the superposition principle no longer holds, and the contributions from individual particles cannot be added linearly. This nonlinearity leads to qualitatively new phenomena in the strong-coupling regime, distinct from classical forces. A more detailed discussion of how the superposition principle differs between classical and quantum forces can be found in Sec.~\ref{subsec:superposition}.

The remaining part of this paper is organized as follows. In Sec.~\ref{sec:sketch}, we present an overview of our method through an intuitive picture of ``a spherical cow in vacuum''. In particular, we clarify why forces arising from quadratically coupled mediators are intrinsically quantum.
Next, we apply this method to the cases of scalar and fermionic mediators in Sec.~\ref{sec:scalar} and Sec.~\ref{sec:fermion}, respectively, demonstrating explicitly how the resulting quantum forces reproduce the known expressions in the weak-coupling limit and derive the correction to that limit. We further analyze their behavior in the strong-coupling regime, where the Born approximation fails and new qualitative features emerge. Broader implications and general properties of strongly coupled quantum forces are discussed in Sec.~\ref{sec:pheno}. We conclude in Sec.~\ref{sec:conclusion}. Technical details are collected in the appendices.

\section{A spherical cow in vacuum: A sketch of our method}
\label{sec:sketch}
In order to develop an intuitive picture of how quantum forces are
generated, let us first briefly review how classical forces, such as
the Coulomb force, are generated.  As is well known in classical electrodynamics, an electrically charged object generates, in its
surrounding area, an electric field $\mathbf{E}=-\mathbf{\nabla}V$,
where $V$ is the electric potential. Any charged particles moving
nearby the object can ``feel'' the electric field, which exerts
a force $\propto \mathbf{E}$ on them. If the
charged object is spherically symmetric, the electric potential $\phi$
outside the object is proportional to $1/r$; hence, the force is proportional 
to $1/r^{2}$.

\begin{figure}[t]
\centering

\includegraphics[width=0.99\textwidth]{Fig/SphericalCow-v2}

\caption{Intuitive pictures of how classical (left panel)
and quantum (right panel) forces are generated. Classical forces such
as the Coulomb force are generated by a static field $\phi$ (dashed
lines) induced by an object (the blue blob labeled ``$\chi$ cluster'') via a coupling of the form $\phi\overline{\chi}\chi$.
With a spherically symmetric profile for the $\chi$ cluster, the
$\phi$ field decreases as $1/r$ outside the sphere, exerting a force
$\propto 1/r^2$ on the test particle (smaller blue blob). Quantum
forces are generated by quantum fluctuations of the $\phi$ field.
Given the interaction $\phi^{2}\overline{\chi}\chi$ indicated on
the right panel, the $\chi$ cluster emits and absorbs $\phi$ simultaneously.
So it cannot generate a static nonzero distribution of $\phi$ (i.e.~$\langle\phi\rangle=0$).
However, quantum fluctuations still cause non-vanishing $\langle\phi^2\rangle$
which decreases as $1/r^{3}$ in this model. The Gauss surface (green dashed) is added to demonstrate that the flux of the corresponding ``electric field'' through the surface is nonzero for the classical case but vanishes in the quantum case. 
\label{fig:cow}}
\end{figure}

This classical picture is illustrated by the left panel of Fig.~\ref{fig:cow},
where, instead of considering a vector mediator, we consider a scalar to avoid unnecessary complications arising from Lorentz indices. (Recall that in classical electrodynamics, $\phi$ is the zeroth component of
the four-potential $A^{\mu}$, not a Lorentz-invariant scalar.) In this figure, $\chi$ denotes matter particles (e.g., protons, electrons), and $\phi$ is a massless mediator, coupled
to $\chi$ via ${\cal L}\supset y\phi\overline{\chi}\chi$. When a
bulk of $\chi$ particles is distributed in the
blue region in Fig.~\ref{fig:cow} ($\chi$ cluster), the equation
of motion (EOM) of $\phi$ is modified as 
\begin{equation}
\left(\partial_{t}^{2}-\nabla^{2}\right)\phi=y\,\langle \overline{\chi} \chi \rangle \;,
\label{eq:phi-EOM-yuk}
\end{equation}
where $\langle \overline{\chi} \chi \rangle$ denotes the ensemble average of the $\chi$ bilinear within the source. 
Throughout this work, the $\chi$ particles inside the source are assumed to be non-relativistic. In this limit, the bilinear simply reduces to the number density  $\langle \overline{\chi}\chi\rangle = n_\chi$. We briefly discuss the relativistic case in Sec.~\ref{subsec:relativistic}.

Following the spirit of ``a spherical cow in vacuum'', we model the source as a spherically symmetric distribution of $\chi$ particles with constant density and radius $R$. Then we can solve the differential equation and obtain
 $\phi\propto1/r$ --- see e.g.~\cite{Smirnov:2019cae}.
Concretely, the density profile is assumed to be
\begin{equation}
n_{\chi}=\begin{cases}
N_{\chi}/V_{R}
& (r<R)\\
0 & (r>R)
\end{cases}\thinspace\ \ \Rightarrow\ \ \phi=\frac{yN_{\chi}}{4\pi}\cdot\frac{1}{r}\quad(\text{for $r> R$})\thinspace,\label{eq:cow}
\end{equation}
where $V_{R}=(4\pi R^{3})/3$ is the volume of the $\chi$ cluster, and $N_{\chi}$ is the total number of $\chi$ particles.
One can also introduce a mass $m_{\phi}$ for the mediator in Eq.~\eqref{eq:phi-EOM-yuk}.
Then, the same methodology leads to the Yukawa potential ($\propto e^{-m_{\phi}r}/r$),
or in the limit of $m_{\phi}\to\infty$, to the MSW potential widely
used in connection to neutrino oscillation phenomena~\cite{Smirnov:2019cae}.

It is important to note that while these classical forces/potentials
(Coulomb, Yukawa, MSW) can all be derived in QFT using $t$-channel
Feynman diagrams and the Born approximation (see, for example, pages 121 and 125 in Ref.~\cite{Peskin:1995ev}),
they essentially arise from the EOM modified by the existence of matter
in space, similar to how the QED Lagrangian gives rise to
Maxwell's equations, which then give rise to the Coulomb force. 

It is therefore tempting to ask whether quantum forces (e.g., neutrino
forces, axion forces) can be derived in a similar way without involving
Feynman diagrams. The answer is yes, and the derivation constitutes the main part of this paper. Our methodology is demonstrated in the right panel of Fig.~\ref{fig:cow}. Here, we consider the interaction to be
\begin{equation}
{\cal L}\supset g\,\phi^{2}J\thinspace,\qquad J \equiv \overline{\chi} \chi\,, \label{eq:L-2}
\end{equation}
where $J$ is the source term and $g$ denotes a generic dimensional coupling. 
Due to the quadratic
form of $\phi$, the long-range force between $\chi$ particles can
be generated only at one-loop level via two-$\phi$ exchange. 
By assigning a hypothetical $Z_{2}$ symmetry to $\phi$, one can immediately see that the $\chi$ cluster can no longer generate  a static nonzero $\phi$ field around it.
At the classical level, $\phi$
waves could be generated dynamically (similar to electromagnetic waves
propagating in vacuum), but the time-averaged value of $\phi$ must
be zero due to the $Z_{2}$ symmetry. At the quantum level, $\phi$ could
exhibit quantum fluctuations, but the ensemble average of $\phi$
must be zero. Both are represented in Fig.~\ref{fig:cow} by $\langle\phi\rangle=0$,
where ``$\langle\ \rangle$'' should  be interpreted as time or
ensemble average at the classical or quantum level, respectively. 

Although $\langle\phi\rangle=0$, the quantity $\langle\phi^{2}\rangle$ need not vanish and generally exhibits a nontrivial dependence on $r$.  
Classically, one may still solve the EOM and obtain wave-like solutions with spatially varying $\phi^{2}(r)$. 
However, because the classical equation,
\begin{align}
\left(\partial_t^{2}-\nabla^{2}-2g\,J\right)\phi = 0\;,
\end{align}
is linear and homogeneous, any classical solution can be multiplied by an arbitrary constant and still remain a valid solution.  
Therefore, the overall normalization of $\phi$, and hence of $\phi^{2}$, is not determined by classical physics alone.  
In particular, in the absence of a classical $\phi$ background, nothing prevents one from rescaling a classical solution so that $\phi^{2}$ becomes arbitrarily small, revealing that classical dynamics alone cannot fix the amplitude of $\langle\phi^{2}\rangle$.

The determination of $\langle\phi^{2}\rangle$ is \emph{fundamentally quantum}.  
Because of the Heisenberg uncertainty principle, $\phi$ and its conjugate momentum $\pi=\dot{\phi}$ cannot both vanish; they must satisfy $[\phi,\pi]\propto i\hbar$.  
This prevents the field amplitude from being scaled to zero and imposes a nonzero, irreducible fluctuation.  
Thus, the scale of $\langle\phi^{2}\rangle$ --- and therefore of the induced quantum force --- is fixed by quantum fluctuations rather than classical dynamics.
 
Once $\langle\phi^{2}\rangle$ is determined, the quantum force follows immediately by placing a test particle $\chi$ in this background and examining how $\langle\phi^{2}\rangle$ modifies its motion. 

In the next two sections, we apply the above method to derive
quantum forces mediated by scalars and fermions. 

\section{Scalar mediator}
\label{sec:scalar}
Let us start with the scalar mediator. For the purpose of illustration, we consider the following toy model throughout this section:
\begin{align}
    \mathcal{L}_S = i \overline{\chi} \gamma^\mu \partial_\mu \chi - m_\chi \bar{\chi} \chi  + \frac{1}{2} (\partial \phi)^2 - \frac{1}{2}m_\phi^2 \phi^2-  \frac{\epsilon}{2\Lambda} \phi^2 J\;, \qquad J \equiv  \overline{\chi}\chi\;, 
    \label{eq:LS}
\end{align}
where $\chi$ is a Dirac fermion with mass $m_\chi$ and $\phi$ is a real scalar with mass $m_\phi$. 
The parameter $\Lambda$ denotes the cutoff scale of the effective interaction, and $\epsilon = \pm 1$ specifies the sign of the coupling. Since $\phi^{2}J$ is real, the coupling must also be real, so $\epsilon$ can take only these two physically distinct values.
We assume that the mediator particle is much lighter than the source particle, $m_\phi \ll m_\chi$.

From the Lagrangian, it is straightforward to derive the EOM:
\begin{align}
\left(\partial^{2}+m_{\phi}^{2}+\frac{\epsilon}{\Lambda}J\right)\phi & =0\thinspace,\label{eq:EOMphi}\\
i\gamma^{\mu}\partial_{\mu}\chi-\left(m_{\chi}+\frac{\epsilon}{2\Lambda}\phi^{2}\right)\chi & =0\thinspace.\label{eq:EOMchi}
\end{align}
It is noteworthy that the EOM of $\phi$ is a linear and homogeneous differential equation, as opposed to Eq.~\eqref{eq:phi-EOM-yuk} derived from the Yukawa interaction. As a consequence, the amplitude of the solution to Eq.~\eqref{eq:EOMphi} cannot be fixed at the classical level, while the solution to Eq.~\eqref{eq:phi-EOM-yuk} has its amplitude determined by the source. As discussed previously, the Heisenberg uncertainty principle dictates that one cannot linearly scale the classical solution down to an arbitrarily small amplitude, and  
the quantum fluctuation of $\phi$ is essentially determined by $[\phi,\dot{\phi}]\propto i\hbar$. 

To compute $\langle \phi^2 \rangle$, we use canonical quantization and promote $\phi$ to a quantum operator $\widehat{\phi}$ and compute $\langle 0| \widehat{\phi}^2 |0 \rangle$. Here we shall clarify the meaning of the vacuum state $|0 \rangle $ in the current context. It is defined as the ground state of the Hamiltonian in the presence of the $\chi$ cluster. Compared to the vacuum state of quantized fields in empty spacetime without any particles, this is a ``modified'' vacuum in the sense that $\langle 0| (\widehat{N}_{\chi}, \widehat{N}_{\phi}) |0 \rangle=(N_\chi, 0)$, where $\widehat{N}_{\chi}$ and $\widehat{N}_{\phi}$ are particle number operators of $\chi$ and $\phi$, respectively. 
In other words, we take $|0\rangle_{N_\chi}$ as the vacuum state, even though it already contains a large number of $\chi$ particles. In what follows, we often suppress the subscript and simply write $|0\rangle$, with the understanding that this state includes the fixed configuration of $\chi$ particles. We nevertheless call it the ``vacuum'' because our interest lies in the dynamics of the mediator field $\phi$ and of any additional $\chi$ particles (such as a test particle) introduced into this system.

After obtaining the static distribution of $\langle \phi^2 \rangle$, one can treat the term $\frac{\epsilon}{2\Lambda}\phi^{2}$ in Eq.~\eqref{eq:EOMchi} on the same footing as $y\phi$ in the Yukawa interaction. 
In the Yukawa case, the term $y\phi$ acts as a position-dependent potential energy once the scalar field profile is fixed. 
Similarly, in the present scenario, replacing $\phi^2$ by its static value $\langle \phi^2 \rangle$ in $\frac{\epsilon}{2\Lambda}\phi^{2}$ produces an effective potential acting on the test particle. This leads to the two-$\phi$ exchange potential
\begin{equation} 
    V_{2\phi}(r) = \frac{\epsilon}{2 \Lambda} \langle \phi^2\rangle\;.
\label{eq:V2phi}    
\end{equation}
This result can also be inferred directly from Eq.~\eqref{eq:EOMchi} for non-relativistic $\chi$ particles: an increase in the effective mass of $\chi$ is physically equivalent to an increase in the potential energy. 
The essential difference from the Yukawa case is that the potential energy now scales as $V\propto \langle\phi^{2}\rangle$ rather than $V\propto \langle \phi \rangle$, although the force continues to be given by $\mathbf{F} = -\nabla V$. 

\subsection{Calculation of $\langle \phi^2 \rangle$}

To compute $ \langle \phi^2 \rangle \equiv \langle 0| \widehat{\phi}^{\,2} |0 \rangle$ with $|0 \rangle$ the modified vacuum as explained above, we shall quantize the field on the background that has already possessed a bulk of non-relativistic $\chi$ particles evenly distributed in the $\chi$ cluster, as shown in Fig.~\ref{fig:cow}. The influence of the $\chi$ cluster on the quantization of the $\phi$ field, as suggested by Eq.~\eqref{eq:EOMphi}, is that it shifts the mass 
\begin{align}
m_\phi^2 \to m_{\rm eff}^2 (r) = m_\phi^2 + m_{\rm M}^2 (r)\;,    
\end{align}
where
\begin{equation}
m_{\rm M}^2(r)\equiv \frac{\epsilon}{\Lambda} n_\chi(r)\;,
\label{eq:mMdef}
\end{equation}
is the matter-induced mass squared due to the presence of $\chi$ particles. An alternative interpretation of the effect is that, due to coherent forward scattering of $\phi$ with particles in the $\chi$ cluster, its dispersion relation is modified. Consequently, when $\phi$ waves propagate through the cluster, they are affected by the medium particles, similar to how glass slows down the propagation of light. 

Due to the spherical symmetry of the system, it is convenient to expand the field $\phi$ in the spherical-wave basis:
\begin{equation}
    \widehat{\phi}(\vecr,t) =  \int {\rm d}\omega \sum_{\ell,m} \frac{u_{\omega \ell}(r)}{r}\left(\widehat{a}_{\omega\ell m}^{} Y_{\ell m}(\theta, \varphi)\,e^{- i \omega t} + \widehat{a}^\dagger_{\omega\ell m}Y^*_{\ell m}(\theta, \varphi)\,e^{i \omega t} \right),\label{eq:quantization-of-phi}
\end{equation}
where $\widehat{a}$ and $\widehat{a}^\dagger$ denote the usual annihilation and creation operators of the field $\phi$, $Y_{\ell m}$ are the standard spherical harmonics with discrete quantum numbers $\ell$ and $m$ denoting the angular momentum and its $z$-component, and $u_{\omega \ell}$ are the (dimensionless) radial wavefunctions. Here, $\omega$ denotes the frequency of the field; since $\phi$ does not form a bound state, there is a continuous spectrum where $\omega$ should be integrated from $m_\phi$ to infinity. The details and conventions of the quantization are given in Appendix~\ref{app:quantization:scalar}.

Plugging (\ref{eq:quantization-of-phi}) into (\ref{eq:EOMphi}), we obtain the radial equation for each frequency and angular momentum:
\begin{equation}
\left[ \partial_r^2 - \frac{\ell(\ell + 1)}{r^2} + \omega^2- \left(m_\phi^2+ m_{\rm M}^2(r)\right)\right] u_{\omega\ell}(r) = 0\;. \label{eq:radialScalar}
\end{equation}
Using the quantized field (\ref{eq:quantization-of-phi}), the commutation relations of creation/annihilation operators, and some properties of the spherical harmonics functions, we obtain, for the fluctuation:
\begin{align}
    \langle \phi^2 \rangle \equiv \langle 0 |\widehat{\phi}^{\,2}|0\rangle  =  
    \frac{1}{r^2} \sum_{\ell} \frac{2 \ell + 1}{4 \pi} \int \frac{{\rm d} \omega}{\omega}\,u^2_{\omega\ell}(r)\;.
    \label{eq:fluctuation}
\end{align}
The remaining task is to solve (\ref{eq:radialScalar}) with proper boundary conditions to obtain solutions for the radial wavefunction.

We consider the case where the $\chi$ particles are uniformly distributed within a sphere of radius $R$,
\begin{equation}
    n_\chi (r) = \begin{cases}
n_\chi &\text{for } r \leq R\;, \\[10pt]
0 & \text{for } r > R\;.
\end{cases} \label{eq:sourcedistribution}
\end{equation} 
In this case, the effective mass squared of the $\phi$ field takes on the constant value:
\begin{equation}
    m_{\rm eff}^2 (r) = 
\begin{cases}
m_\phi^2 + \epsilon \frac{n_\chi}{\Lambda} &\text{for } r \leq R \;,\\[10pt]
m^2_\phi & \text{for } r > R\;.
\end{cases}
\end{equation} 
The corresponding momenta for the region inside and outside the sphere are given as
\begin{align}
k_\text{in} = \sqrt{\omega^2-m^2_\phi- m_{\rm M}^2}\;,\qquad
k_\text{out} = \sqrt{\omega^2-m^2_\phi}\;.\label{eq:kinkout}
\end{align}

The general solution to (\ref{eq:radialScalar}) with a constant momentum $k$ is a linear combination of the spherical Bessel functions of the first kind $j_\ell(kr)$ and second kind $y_\ell(kr)$. Matching the interior and exterior solutions at the boundary $r=R$ by imposing continuity and differentiability of the field for each frequency and angular momentum, the radial solution is determined up to an overall normalization constant. This remaining constant can only be fixed by the completeness relation that follows from the canonical quantization of the field (see Appendix \ref{app:quantization:scalar}). This highlights the quantum nature of the quantum force. 

Carrying out these steps, as presented in Appendix~\ref{app:radial-scalar}, the full expression for the radial solutions reads:
\begin{align}
    u_{\omega\ell}(r) = r\sqrt{\frac{\omega k_\text{out}}{\pi}}
    \begin{cases}
    \dfrac{j_\ell(k_\text{in}r)}{j_\ell(k_\text{in}R)} \left[j_\ell(k_\text{out} R) \cos \delta_\ell + y_\ell(k_\text{out} R) \sin \delta_\ell \right] \quad   \text{for $r\leq R$}\;,\\[20pt]
    j_\ell(k_{\text{out}} r) \cos \delta_\ell +  y_\ell(k_{\text{out}} r) \sin \delta_\ell \qquad\qquad\qquad \;\text{for $r>R$}\;, \\
    \end{cases}
    \label{eq:radialscalarsolution}
\end{align}
where $\delta_\ell$, the phase shift, is given by
\begin{align}
\tan \delta_\ell = \frac{
\displaystyle \frac{k_{\text{in}}}{k_{\text{out}}} j_\ell'(k_{\text{in}} R) j_\ell(k_{\text{out}} R)
- j_\ell(k_{\text{in}} R) j_\ell'(k_{\text{out}} R)
}{
\displaystyle j_\ell(k_{\text{in}} R) y_\ell'(k_{\text{out}} R)
- \frac{k_{\text{in}}}{k_{\text{out}}} j_\ell'(k_{\text{in}} R) y_\ell(k_{\text{out}} R)
}\;.  \label{eq:phaseshift}
\end{align}

Next, we turn to the evaluation of $\langle \phi^2 \rangle$. As it is well known, the vacuum fluctuation is ultraviolet (UV) divergent due to the contribution of high-frequency modes. However, the \emph{difference} induced by changing the geometric configuration, that is introducing or removing the source, is finite and physically observable. Consequently, we should compute the change in $\langle \phi^2 \rangle$ in the presence of the source relative to pure vacuum. This quantity isolates the effect of the source on the mediator field fluctuations:
\begin{equation}
    \Delta\langle \phi^2 \rangle \equiv \langle\phi^2 \rangle_{n_\chi \neq 0} - \langle\phi^2 \rangle_{n_\chi= 0}\;.
    \label{eq:differencescalar_General}
\end{equation}
Each term in (\ref{eq:differencescalar_General}) is divergent, but their difference $\Delta\langle \phi^2 \rangle$ is finite and is the only quantity that contributes to the quantum force in (\ref{eq:V2phi}).  
Note that setting $n_\chi = 0$ is equivalent to taking $k_{\rm in} = k_{\rm out}$, which implies a vanishing phase shift according to (\ref{eq:phaseshift}).

Combining (\ref{eq:fluctuation}), (\ref{eq:radialscalarsolution}), and (\ref{eq:differencescalar_General}), we obtain 
\begin{align}
\Delta \langle \phi^2 \rangle &=  \frac{1}{\pi}
\sum_{\ell} \dfrac{2\ell + 1}{4\pi} \int_{m_\phi}^{\infty}  {\rm d}\omega\, k_\text{out}\nonumber\\
&\times
\begin{cases}
\dfrac{j_\ell^2(k_\text{in}r)}{j_\ell^2(k_\text{in}R)} \left[j_\ell(k_\text{out} R) \cos \delta_\ell + y_\ell(k_\text{out} R) \sin \delta_\ell \right]^2 - j_\ell^2(k_\text{out} r) \quad 
 \text{for $r\leq R$}\;,\\[2pt] \\
\left[ j_\ell(k_\text{out} r) \cos \delta_\ell + y_\ell(k_\text{out} r) \sin \delta_\ell \right]^2
- j_\ell^2(k_\text{out}r) \quad \quad\quad\quad\quad\;\;\,
\text{for $r> R$}\;.
\end{cases}
\label{eq:differencescalar}   
\end{align}

So far, all expressions have been exact. The fluctuation in (\ref{eq:differencescalar}) is written as a sum over all modes. However, the quantum force depends only on the IR sector of the theory: modes with frequencies much larger than $R^{-1}$ cannot ``feel'' the presence of the source and therefore do not contribute to the observable effect. 
Since we care only about the large-$r$ behavior, we can treat $R$ small relevant to $r \sim k_{\rm out}^{-1}$ and expand all quantities in powers of $k_{\rm out} R$.
From the phase shift in (\ref{eq:phaseshift}), we find 
\begin{align}
\tan \delta_\ell \sim {\cal O}\!\left( (k_{\rm out} R)^{2\ell + 1} \right).    
\end{align}
This scaling corresponds to the expansion of the potential (\ref{eq:V2phi}) in position space,
\begin{align}
V_{2\phi}(r) 
= V_0(r)\left( 1 + \sum_{\ell=1}^\infty 
\frac{c_{2\ell} R^{2\ell}}{r^{2\ell}} \right),
\end{align}
where $V_0$ denotes the contribution from the $\ell = 0$ mode and $c_{2\ell}$ are dimensionless coefficients. This implies that in the  far-field limit $r \gg R$, only the $\ell = 0$ mode needs to be retained. Contributions from higher-$\ell$ modes are subdominant, and they are discussed in Appendix~\ref{app:higherell}. 

For the $\ell = 0$ mode, the spherical Bessel functions take particularly simple forms: 
$ j_0(x) = \sin x / x $ and $ y_0(x) = -\cos x / x $. 
To simplify the phase shift in (\ref{eq:phaseshift}), we express $k_{\rm in} R$ in terms of $k_{\rm out} R$ and $m_{\rm M} R$ using the relation $k_{\rm in}^2 = k_{\rm out}^2 - m_{\rm M}^2$. 
We then expand the phase shift in the limit $k_{\rm out} R \ll 1$ while keeping $m_{\rm M} R$ arbitrary:
\begin{align}
    \tan \delta_0 &= k_{\rm out}R \left[1-\frac{\tanh\left(m_{\rm M}R\right)}{m_{\rm M}R}\right] + \mathcal{O}\left((k_\text{out} R)^3\right).\label{eq:tandelta0}   
\end{align}

We now define the scalar-mediator form factor:
\begin{align}
F_\phi(z) \equiv \epsilon\,\frac{3}{z^2} \left(1-\frac{\tanh z}{z}\right),\qquad
z\equiv m_{\rm M}R\;.\label{eq:formfactorscalardef}
\end{align}
Retaining only the leading order of $k_{\rm out}R$ and using $n_\chi = N_\chi/(4\pi R^3/3)$ with $N_\chi$ the total number of $\chi$ particles inside the source, the phase shift becomes
\begin{align}
\tan \delta_0 = \epsilon N_\chi\frac{k_{\rm out}}{4 \pi \Lambda} \,F_\phi(m_{\rm M} R)\;. \label{eq:phaseShift0}
\end{align}
Note that $m_{\rm M}$ is real for $\epsilon = +1$ and purely imaginary for $\epsilon = -1$. Using the identity $\tanh(i|z|)/(i|z|) = \tan(|z|)/|z|$, Eq.~(\ref{eq:formfactorscalardef}) can be rewritten as\footnote{We note that the function in (\ref{eq:formfactorscalar}) was first derived in \cite{Hees:2018fpg} in a different context involving a classical $\phi$ background; see also \cite{Banerjee:2022sqg,Day:2023mkb,Bauer:2024yow,Banerjee:2025dlo,delCastillo:2025rbr,Gan:2025nlu} for related effects in the detection of ultralight scalar dark matter.
}
\begin{equation} 
F_\phi(z) = \frac{3}{|z|^2}
\begin{cases}
        1- \dfrac{\tanh |z|}{|z|}  &\text{for } \epsilon = +1\;, \\[10pt]
        \dfrac{\tan |z|}{|z|} -1 &\text{for }   \epsilon = - 1\;. \\ 
    \end{cases}\label{eq:formfactorscalar}
\end{equation}
 In the limit $|z| \ll 1$, we obtain $F_\phi(z) \to 1$ for both $\epsilon = \pm 1$.

When $\delta_0$ is small, we can further simplify. ($\delta_0$ is small in most of the parameter space, see Sec.~\ref{subsec:strongcouplingscalar} for further discussion.)
For the case of small $\delta_0$, we use
$\tan \delta_0 \approx \delta_0$.
Substituting this approximation into (\ref{eq:differencescalar}) yields, for $r > R$,
\begin{align}
\Delta \langle \phi^2 \rangle &=  -\frac{1}{4\pi^2 r^2} \int_{0}^\infty \frac{{\rm d} k_{\rm out}}{\sqrt{k_{\rm out}^2+m_\phi^2}}\,\sin(2k_{\rm out}r)\,\delta_0 + {\cal O}\left(\delta_0^2\right)\nonumber\\
& = -\frac{\epsilon N_\chi}{16\pi^3 \Lambda r^2} F_\phi(m_{\rm M}R)\int_{0}^\infty {\rm d} k_{\rm out}\frac{k_{\rm out}}{\sqrt{k^2_{\rm out}+m_\phi^2}}\,\sin(2k_{\rm out}r)\;, \label{eq:phisqoutside}
\end{align}
where we have changed the integration variable from $\omega$ to $k_{\rm out}$ using (\ref{eq:kinkout}). 

The integral in the second line of (\ref{eq:phisqoutside}) does not converge as $k_{\rm out}\to \infty$. However, this divergence is unphysical: UV modes with frequencies much larger than $1/R$ do not contribute to the observable, as they cancel out in the difference defined in (\ref{eq:differencescalar_General}). As we show in Appendix~\ref{app:contour}, this spurious divergence can be removed cleanly and in a regulator-independent manner by performing an analytic continuation in the complex plane, leading to a finite contribution to the quantum force. The result is given by
\begin{align}
I = \frac{1}{2}\, \Im \left[\int_{-\infty}^\infty {\rm d} k_{\rm out}\frac{k_{\rm out}}{\sqrt{k^2_{\rm out}+m_\phi^2}}\,e^{i\,2k_{\rm out}r}\right] = m_\phi K_1(2m_\phi r)\;,   
\end{align}
where $K_n$ is the modified Bessel function of the second kind. Therefore, we obtain the final expression for the field fluctuation outside the source:
\begin{align}
\boxed{
\Delta \langle \phi^2 \rangle = - \frac{\epsilon N_\chi m_\phi}{16\pi^3 \Lambda} \times \frac{K_1(2m_\phi r)}{r^2}\times F_\phi(m_{\rm M}R)\;. \label{eq:phisqoutsidefinal} } 
\end{align}
The result in (\ref{eq:phisqoutsidefinal}) is the leading-order result in the limit of $r\gg R$, where we can ignore the contributions from higher-$\ell$ modes. The expression in (\ref{eq:phisqoutsidefinal}) exhibits a factorized structure --- see Sec.~\ref{subsec:factorization} for a general discussion of this factorization. 

So far, we have not specified the value of $m_{\rm M}R$, which is determined by the coupling strength between $\chi$ and $\phi$. To see it more clearly, we can write
\begin{align}
m_{\rm M}^2 R^2 =\epsilon \frac{n_\chi R^2}{\Lambda} = \frac{3\epsilon}{4\pi} \frac{N_\chi}{\Lambda R}\;.    
\label{eq:mR}
\end{align}
Two qualitatively different scenarios are in order:
\begin{itemize}
    \item \textbf{Weak-coupling limit:} $N_\chi/\Lambda \ll R \;\Rightarrow\; |m_{\rm M}|R\ll 1$. In this regime, our result reduces to that from the Feynman diagram approach at leading order, with small corrections suppressed by powers of $m_{\rm M}^2 R^2$.
    \item \textbf{Strong-coupling limit:} $N_\chi/\Lambda \gg R \;\Rightarrow\; |m_{\rm M}| R \gg 1$. This regime can naturally arise in dense environments (such as neutron stars), where the number density of $\chi$ particles is extremely large. In such cases, the matter-induced modification of the mediator field fluctuation becomes substantial, leading to a nonlinear violation of the superposition principle (see Sec.~\ref{subsec:superposition}). Moreover, from the perspective of scattering theory, this corresponds to a breakdown of the Born approximation~\cite{Sakurai:2011zz}, signaling that the standard perturbative Feynman diagram approach is no longer valid.
\end{itemize}

In what follows, we analyze these two regimes separately. In the weak-coupling limit, the sign of the coupling plays no essential role. 
In contrast, in the strong-coupling regime, the sign becomes crucial, and we therefore treat the cases $\epsilon=\pm1$ separately. 
Roughly speaking, relative to the weak-coupling limit, the field fluctuation $\langle \phi^{2}\rangle$ is suppressed for $\epsilon=+1$ and enhanced for $\epsilon=-1$.

\subsection{Weak-coupling limit}
In the weak-coupling limit, $|z|=|m_{\rm M}| R  \ll 1$, we can expand the form factor in Eq.~(\ref{eq:formfactorscalar}) using 
\beq
\frac{\tan |z|}{|z|} = 1+ \frac{|z|^2}{3} + \frac{2|z|^4}{15}  + {\cal O}(|z|^6)\;, \qquad
\frac{\tanh |z|}{|z|} = 1- \frac{|z|^2}{3}  + \frac{2|z|^4}{15} + {\cal O}(|z|^6)\;,
\eeq
from which we obtain
\begin{align}
F_\phi(z) = 1 - \epsilon\,\frac{2 |z|^2}{5} + {\cal O}(|z|^4)\;.\label{eq:Fphiweak}   
\end{align} 
Substituting (\ref{eq:phisqoutsidefinal}) and (\ref{eq:Fphiweak}) into (\ref{eq:V2phi}), we obtain
 \begin{equation} 
     V_{2\phi}(r) = - \frac{N_\chi m_\phi}{32\pi^3 \Lambda^2} \frac{ K_1( 2 m_\phi r)} {r^2}\times \left(1-\epsilon\, \frac{2|z|^2}{5} + {\cal O}(|z|^4)\right).\label{eq:V2phiweak}
 \end{equation}
 
At leading order, all dependence on $\epsilon$ and $R$ cancels out. In the single-particle limit $N_\chi \to 1$, the leading term reproduces the standard diagrammatic result exactly (see, e.g., \cite{Fichet:2017bng}). In the massless limit $m_\phi r \ll 1$, using $K_n(x \ll 1) \sim x^{-n}$, the expression further reduces to the familiar $1/r^3$ behavior.

The next-to-leading term $2|z|^2/5$ is the correction to the Feynman diagram result. It shows that the quantum force is suppressed for $\epsilon = +1$ and enhanced for $\epsilon=-1$ compared to the limit of the Born approximation.

\subsection{Strong-coupling limit: $\epsilon = +1$}
\label{subsec:strongcouplingscalar}
For $\epsilon = +1$, $z\equiv m_{\rm M}R$ is real, and we can take the $z \gg 1$ limit of (\ref{eq:formfactorscalar}), 
\begin{align}
F_\phi(z \gg 1) = \frac{3}{z^2} \qquad (\text{for $\epsilon = +1$})\;.
\end{align}
Consequently, the quantum force is suppressed by a factor of $3/(m_{\rm M} R)^2$ relative to the weak-coupling result in (\ref{eq:V2phiweak}).

We refer to the suppression factor $3/(m_{\rm M}R)^2$ as screening effect. 
A quantitative explanation of it is given in Sec.~\ref{eq:subsec:screening}. 
Here, we give a qualitative picture of it: in the strong repulsive regime, the momentum inside the source is purely imaginary, reflecting the presence of a large effective mass barrier. Indeed, using (\ref{eq:kinkout}) we have
\begin{align} 
k_{\rm in}^{2} = k_{\rm out}^{2} - m_{\rm M}^{2} < 0\;, 
\end{align} 
so that the interior solution is evanescent. Thus, the fluctuations generated inside the source cannot propagate to the exterior, leading to a suppressed quantum force in the IR regime $r \gg R$.

\subsection{Strong-coupling limit: $\epsilon=-1$}

\begin{figure}[t]
\centering
    \begin{subfigure}{0.48\textwidth}
        \centering
        \includegraphics[width=\linewidth]{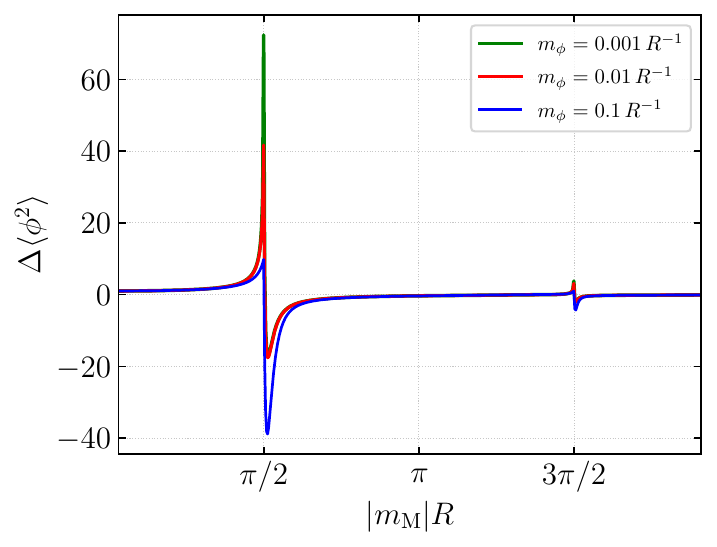}
    \end{subfigure}
    \hfill
    \begin{subfigure}{0.48\textwidth}
        \centering
        \includegraphics[width=\linewidth]{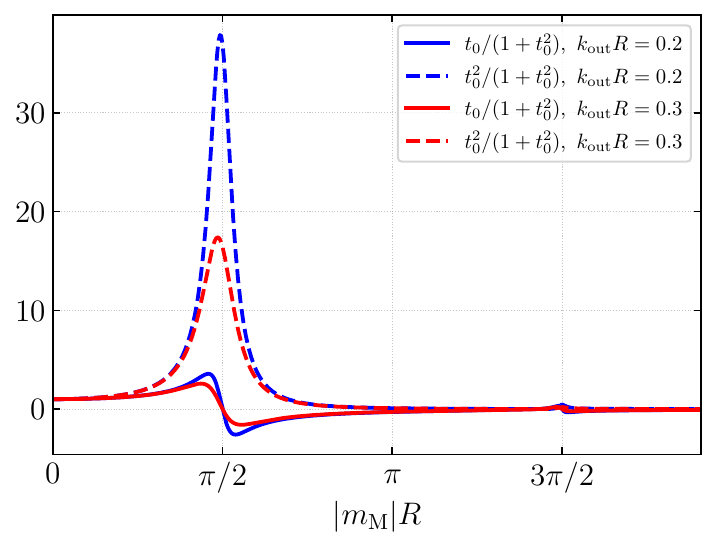}
    \end{subfigure}
\caption{\label{fig:peaks}
The field fluctuation (left panel) as a function of $|m_{\rm M}|R$, computed from Eq.~(\ref{eq:phisqnonlinear}). For comparison, the two individual terms appearing in Eq.~(\ref{eq:phisqnonlinear}) are shown in the right panel. For convenience, all quantities are normalized to unity in the weak-coupling limit ($z\ll 1$).
Numerical results are evaluated at $r = 10R$.}
\end{figure}

We have seen that for $\epsilon = +1$, the quantum force in the strong-coupling regime exhibits the same $r$-scaling behavior as in the weak-coupling regime, up to an overall suppression factor. This arises because the approximation $\tan\delta_0 \approx \delta_0$ remains valid for arbitrarily large $m_{\rm M}R$ when the coupling is positive. 

The situation is qualitatively different for negative coupling, $\epsilon = -1$. In fact, the result is sensitive to the UV-complete model that generates such a coupling. We discuss this point in more detail below in Sec.~\ref{sec:tach}, and for now, we simply assume that the effective theory is well behaved.

Under this assumption, our case corresponds to $m_{\rm M}$ being purely imaginary, and the form factor in Eq.~(\ref{eq:formfactorscalar}) diverges whenever $\lvert m_{\rm M}\rvert R = \pi/2 + n\pi$ (with $n$ an arbitrary integer). In these regions, the phase shift in Eq.~(\ref{eq:phaseShift0}) is no longer small, and therefore the small-$\delta_0$ expansion used in Eq.~(\ref{eq:phisqoutside}) breaks down.

To account for this nonlinear behavior, we return to the general expression in (\ref{eq:differencescalar}). 
For $\ell = 0$, the fluctuation depends on $\sin(2\delta_0)$ and $\cos(2\delta_0)$, which can be rewritten in terms of $\tan\delta_0$. 
For $r > R$, this yields
 \begin{align}
 \Delta\langle \phi^2 \rangle 
 = - \frac{1}{4 \pi^2 r^2} \int_{0}^\infty \frac{{\rm d} k_\text{out}}{\sqrt{k_\text{out}^2 + m_\phi^2}} \left[ \sin( 2 k_\text{out} r)\,\frac{t_0}{1+ t_0^2} - \cos( 2 k_\text{out} r)\,\frac{t_0^2}{1+ t_0^2} \right],\label{eq:phisqnonlinear}
 \end{align}
where
\begin{align}
     t_0\equiv \tan\delta_0 = k_{\rm out}R \left(1-\frac{\tan |z|}{|z|}\right),\qquad
     z\equiv m_{\rm M}R\;.
\end{align}

It is straightforward to verify that (\ref{eq:phisqnonlinear}) reproduces the weak-coupling result (\ref{eq:phisqoutside}) in the limit of $\delta_0 \ll 1$. 
In Fig.~\ref{fig:peaks}, we show $\Delta\langle\phi^2\rangle$ as a function of $|m_{\rm M}|R$ obtained from performing the numerical integral of (\ref{eq:phisqnonlinear}).

At the resonant values $|z| = \pi/2 + n\pi$, although the phase shift $\tan\delta_0$ is divergent, the integral in (\ref{eq:phisqnonlinear}) remains finite and can be evaluated analytically, leading to the following result:
\begin{align}
\Delta \langle\phi^2\rangle \Big|_{|z|=\pi/2 +n\pi} &= \frac{1}{4\pi^2 r^2}K_0(2m_\phi r)\label{eq:fluctuation-at-peaks}\\
&\xrightarrow{m_\phi r\ll 1} \frac{1}{4\pi r^2} \left[-\gamma_{\rm E} - \log(m_\phi r) + {\cal O}\left(m_\phi^2 r^2\right)\right], \label{eq:fluctuation-at-peaks-small-mass}    
\end{align}
where $\gamma_{\rm E} \approx 0.577$ is Euler's constant.
Thus, at the peaks and for $m_\phi r \ll 1$, the potential scales as 
$V_{2\phi}(r) \sim \log(m_\phi r)/r^2$, in contrast with the $1/r^3$ scaling in the weak-coupling regime. Comparing (\ref{eq:fluctuation-at-peaks}) with the weak-coupling result ($|z|\ll 1$), we find
 \begin{align}\label{eq:resonanceRatio}
 \frac{\Delta \langle\phi^2\rangle \Big|_{|z|=\pi/2 +n\pi}}{\Delta \langle\phi^2\rangle \Big|_{|z| \ll 1}} &= \frac{12}{\pi^2\left(2n+1\right)^2}\, \frac{K_0(2m_\phi r)}{\left(m_\phi R\right) K_1(2m_\phi r)}\\
& \xrightarrow{m_\phi r\ll 1}  \frac{24 }{\pi^2 \left(2n+1\right)^2}\frac{r}{R}\left[\log\left(\frac{1}{m_\phi r}\right)-\gamma_{\rm E} + {\cal O}\left(m_\phi^2 r^2\right)
 \right],
 \end{align}
Thus, in the limit $m_\phi r \ll 1$, the fluctuation at the resonant peaks is enhanced relative to the weak-coupling result by a factor of order 
$\sim (r/R)\, \log(m_\phi r)$.

An interesting feature of the resonant result in Eq.~(\ref{eq:fluctuation-at-peaks}) is that it contains no dependence on the cutoff scale $\Lambda$ or on the microscopic parameters of the $\chi$-$\phi$ interaction. This reflects the fact that, at the resonant values $|m_{\rm M}|R = \pi/2 + n\pi$, the $s$-wave phase shift satisfies $\tan\delta_0 \to \pm\infty$, but the fluctuation depends only on the universal combinations $\tan\delta_0/(1+\tan^2\delta_0)$ and $\tan^2\delta_0/(1+\tan^2\delta_0)$, see Eq.~(\ref{eq:phisqnonlinear}), both of which have finite limits. Consequently, the detailed short-distance structure of the $\chi$ medium drops out, leaving a result controlled entirely by IR physics.
This leads to a \emph{non-decoupling} effect: even if both the cutoff scale $\Lambda$ and the number density $n_\chi$ are taken to infinity while keeping their ratio $m_{\rm M}^2 \sim n_\chi/\Lambda$ fixed (so that the resonance condition remains satisfied), the resulting field fluctuation remains finite and non-vanishing. In other words, once the system is tuned to the resonance, the quantum force becomes insensitive to the UV completion of the $\chi$-$\phi$ interaction and is governed solely by the long-distance dynamics of the light mediator.

The origin of the discrete peak structure can be understood from the bound-state spectrum of $\phi$ inside a finite well whose depth is controlled by $m_{\rm M}^2$. For $\ell=0$, the radial equation~(\ref{eq:radialScalar}) admits bound states whenever matching at $r=R$ yields the transcendental condition
\begin{align}
    k_{\rm in}\cot(k_{\rm in} R) = -q\;, 
    \qquad 
    q \equiv \sqrt{m_\phi^2 - \omega^2}\;,
\end{align}
where $q>0$ (i.e.\ $\omega<m_\phi$) governs the exponential decay of the wavefunction outside the source. The zero-energy bound state corresponds to the limit $\omega\to m_\phi^{-}$, where $q\to 0^{+}$, and in this limit the bound-state condition reduces to
\begin{align}
    k_{\rm in}R = \frac{\pi}{2} + n\pi\;, \qquad n=0,1,2,\cdots\,.
\end{align}
Since $k_{\rm in}\to |m_{\rm M}|$ as $\omega\to m_\phi$, the zero-energy bound states occur precisely at
\begin{align}
    |m_{\rm M}|R = \frac{\pi}{2} + n\pi\,,
\end{align}
in one-to-one correspondence with the peaks of the fluctuation. In scattering theory, the appearance of a zero-energy bound state (or threshold state) is known to produce a resonant enhancement of the low-energy scattering amplitude, reflected in a large phase shift and scattering length.\footnote{This is analogous to the Sommerfeld enhancement, where resonance occurs when the (positive) scattering energy nearly matches the (negative) binding energy of a would-be bound state, so that two non-relativistic particles can momentarily form a zero-binding-energy state~\cite{Ferrante:2025lbs}.} Because the fluctuation $\langle\phi^2\rangle$ is obtained by integrating over all scattering modes, it is therefore enhanced whenever a zero-energy bound state exists, i.e., at precisely the values of $|m_{\rm M}|R$ where the peaks appear.

\subsubsection{Tachyonic instability} \label{sec:tach}
The above bound-state analysis has an important caveat: it is performed at the level of a fixed quantum-mechanical potential well. 
In particular, we have treated the interior region as a static well and neglected the back-reaction of the mediator field on the source profile. 
This approximation captures the spectrum of the Klein-Gordon operator in a prescribed well, but it does not include the response of the medium when the effective mass squared becomes negative. Indeed, the first resonance condition $|m_{\rm M}|R = \pi/2$ already implies $|m_{\rm M}|>m_\phi$ in the long-range regime $m_\phi R\ll 1$. In this case, the effective mass squared of $\phi$ inside the source becomes tachyonic, and the mediator field is expected to develop a nonzero classical profile. The resulting back-reaction can significantly modify the $\chi$ mass and invalidate the assumptions of a static, non-relativistic $\chi$ background used in the bound-state treatment. 
A consistent analysis of this regime requires incorporating the nonlinear dynamics of $\phi$ and its feedback on the medium.

Once higher-order terms are included to stabilize the potential and the back-reaction of the classical $\phi$ field is properly taken into account, the behavior near the resonance values $|m_{\rm M}|R=\pi/2+n\pi$ may be qualitatively altered. For example, consider a $\lambda\phi^{4}$ self-interaction term. 
For $|m_{\rm M}|^{2}>m_\phi^{2}$, the $\phi$ mass term inside the source becomes tachyonic, and the mediator field rolls down to a new vacuum and develops a nonzero classical profile $\phi_{\rm cl}(r)$ that interpolates smoothly from the interior to $\phi=0$ outside. 
Fluctuations around this symmetry-breaking vacuum propagate with a strictly \emph{positive} effective mass squared,
$m_{\rm eff}^2 = V''(\phi_{\rm cl}) = 2(|m_{\rm M}|^{2}-m_\phi^{2}) > 0$,
so the interior region acts as a repulsive mass barrier rather than as a potential well. Consequently, the interior no longer supports the zero-energy bound states responsible for the divergent phase shift found in the quadratic theory, and the sharp spike structure disappears once the nonlinear dynamics are included.
In the stabilized theory with $\lambda \phi^4$ term, the fluctuation modes are exponentially suppressed inside the source, and the net effect is a screening of $\Delta\langle\phi^{2}\rangle$ outside, rather than a resonant enhancement. A detailed treatment of this nonlinear regime lies beyond the scope of this work. 

\section{Fermionic mediator}
\label{sec:fermion}
We now turn to the case where the quantum force is mediated by fermions. 
The overall strategy closely parallels the scalar analysis in the last section, so we highlight only the key differences.
For concreteness, we consider the toy model
\begin{equation}
    \mathcal{L}_F 
    = i \overline{\chi}\gamma^\mu\partial_\mu\chi - m_\chi \overline{\chi}\chi
      + i \overline{\psi}\gamma^\mu\partial_\mu\psi - m_\psi \overline{\psi}\psi
      - \frac{\epsilon}{\Lambda^2}\, \overline{\psi}\psi\, J\;, \qquad J \equiv \overline{\chi}\chi\;,
    \label{eq:LF}
\end{equation}
with $\epsilon=\pm 1$, and where $\chi$ and $\psi$ are Dirac fermions of masses $m_\chi$ and $m_\psi$. Note that our toy model is not the one we have for the SM neutrinos that couple with a $V-A$ interaction. In contrast to the scalar case, the sign of the coupling, while still physically meaningful, does not play a crucial role in determining the behavior of the quantum force, as we will demonstrate below.

The EOMs are
\begin{align}
    i\gamma^\mu\partial_\mu\psi 
    - \left(m_\psi + \frac{\epsilon}{\Lambda^2}J\right)\psi &= 0\;,
    \label{eq:EOMpsi} \\
    i\gamma^\mu\partial_\mu\chi
    - \left(m_\chi + \frac{\epsilon}{\Lambda^2}\overline{\psi}\psi\right)\chi &= 0\;.
    \label{eq:EOMchi-fermion}
\end{align}
Our goal is to compute the induced fluctuation $\langle\overline{\psi}\psi\rangle$ at a distance $r$ from a static source of $\chi$ particles. 
Once this quantity is determined, the effective potential felt by a test $\chi$ particle at distance $r$ is simply the two-$\psi$ exchange potential:
\begin{equation}
    V_{2\psi}(r) = \frac{\epsilon}{\Lambda^2}\,\langle \overline{\psi}\psi \rangle\;.\label{eq:V2psi}
\end{equation}

\subsection{Calculation of $\langle \bar{\psi}\psi \rangle$}
Suppose the source is static, unpolarized, and isotropic, with number density $n_\chi(r)$. 
Inside the source, the effective fermion mass is shifted to 
\begin{align}
m_\psi \to m_{\rm eff}(r) = \left|m_\psi + m_{\rm M}(r)\right|,
\label{eq:meffdef-fermion} 
\end{align}
where
\begin{equation}
    m_{\rm M}(r) \equiv \frac{\epsilon}{\Lambda^{2}}\, n_\chi(r)\label{eq:mMdef-fermion}
\end{equation}
is the induced mass generated by the $\chi$ distribution. 

A key distinction from the scalar case is that for fermions the mass term enters \emph{linearly} in the Dirac equation, so the bare mass and the induced mass simply add.  
In contrast, for scalars, the mass term appears quadratically in the Klein-Gordon equation, so it is the \emph{mass squared} that receives additive contributions.
Thus, unlike scalars, a fermion mass never becomes tachyonic, and the sign of $m_{\rm eff}$ has no physical consequence, as it can always be absorbed by a chiral rotation of the fermion field.

As in the scalar case, the spherical symmetry of the system allows us to expand the fermion field $\psi$ in spherical-wave modes. 
Because spin is now involved, the relevant quantum numbers are those of the total angular momentum, which combines orbital and spin degrees of freedom. 
We label the states by $(j,m)$ for the total angular momentum and its $z$ component; similarly, $(\ell,m_\ell)$ for the orbital angular momentum and $(s,m_s)$ for the spin. (In our case, $s=1/2$ is fixed.)
The corresponding angular eigenfunctions are the spinor spherical harmonics, $\Omega_{j\ell m}(\hat{\mathbf{r}})$, defined as
\begin{align}
\Omega_{j\ell m}(\hat{\mathbf{r}})
  = \sum_{m_\ell,m_s}
  \langle \ell m_\ell;\tfrac12\, m_s \mid j\,m\rangle\,
  Y_{\ell m_\ell}(\hat{\mathbf r})\,\chi_{\frac12 m_s}\;,\label{eq:spinorsphericalharmoinicsdef}
\end{align}
where the constant spinors $\chi_{\frac{1}{2} m_s}$ read
\begin{align}
    \chi_{\frac{1}{2} \frac{1}{2}} = \begin{pmatrix}
      1 \\ 0  
    \end{pmatrix}, \quad \chi_{\frac{1}{2} -\frac{1}{2}} = \begin{pmatrix}
      0 \\ 1
    \end{pmatrix}.
\end{align}

With the spinor spherical harmonics as building blocks, the spatial wavefunction of a Dirac fermion can be written as (see, e.g., Chapter 9 of \cite{Greiner1990RQM}):
\begin{align}
    \psi_{\omega j \ell m} (\mathbf{r}) = \frac{i}{r} \begin{pmatrix}
    f_{\omega j \ell}(r)\,\Omega_{j \ell m}(\hat{\mathbf r}) \\[4pt]
    i\,g_{\omega j \ell}(r)\,\Omega_{j \ell' m}(\hat{\mathbf r})
  \end{pmatrix}, \quad \ell' \equiv 2j -\ell\;,\label{eq:psiwavefunction}
\end{align}
where $\omega$ denotes the frequency of the field as in the scalar case, and $f(r)$ and $g(r)$ are radial profile functions.

For a given total angular momentum $j$, the orbital angular momentum can take the two values, $\ell = j\pm 1/2$, which necessarily have opposite parity. 
These are denoted by $\ell$ and $\ell'$ in Eq.~(\ref{eq:psiwavefunction}). Each choice corresponds to the same $j$ but yields an independent basis state. 
For example, for $j=1/2$ the two possibilities are $(\ell,\ell') = (0,1)$ and $(1,0)$.
It is convenient to introduce the Dirac quantum number
\begin{align}
|\kappa| \equiv j+1/2 \;,    
\end{align}
with the convention that $\kappa = -(j+1/2)$ corresponds to $\ell = j - 1/2$, while $\kappa = +(j+1/2)$ corresponds to $\ell = j + 1/2$. If a given value of $\kappa$ maps to the pair $(j,\ell)$, then $-\kappa$ maps to the pair $(j,\ell')$.
Using these conventions, Eq.~(\ref{eq:psiwavefunction}) can be rewritten more compactly as
\begin{align}
    \psi_{\omega \kappa m} (\mathbf{r}) = \frac{i}{r} \begin{pmatrix}
    f_{\omega \kappa}(r)\,\Omega_{\kappa m}(\hat{\mathbf r}) \\[4pt]
    i\,g_{\omega \kappa}(r)\,\Omega_{-\kappa m}(\hat{\mathbf r})
  \end{pmatrix},\label{eq:psiwavefunction2}
\end{align}
where the radial profile $f_{\omega \kappa}(r)$ and $g_{\omega \kappa}(r)$ satisfy the coupled first-order equations
\begin{align}\label{eq:radialFermion}
    \frac{\rm d}{{\rm d}r} \begin{pmatrix}
       f_{\omega \kappa} (r) \\  g_{\omega \kappa}(r)
    \end{pmatrix} = 
    \begin{pmatrix}
        -\kappa/r & \omega+(m_\psi + m_{\rm M}) \\
        -\omega+(m_\psi + m_{\rm M}) & \kappa/r
    \end{pmatrix}
    \begin{pmatrix}
         f_{\omega \kappa} (r) \\  g_{\omega \kappa}(r)
    \end{pmatrix}.
\end{align}

Next, we quantize the Dirac field $\widehat\psi$. Using the spatial mode functions in Eq.~(\ref{eq:psiwavefunction2}), the field operator can be expanded as
\begin{align}
    \widehat{\psi}(\mathbf{r}, t) = \sum_{\kappa=\pm1,\pm2 \cdots}\, \sum_{m=-j}^j\int {\rm d} \omega \left( \widehat{b}_{\omega\kappa m}\psi_{\omega \kappa m} (\mathbf{r})\,e^{-i \omega t}  + \widehat{d}^\dagger_{\omega\kappa m} \psi^c_{\omega \kappa m}(\mathbf{r})\,e^{i \omega t}  \right),
    \label{eq:quantization-of-psi}
\end{align}
where $\widehat{b}_{\omega\kappa m}$ annihilates a fermion $\psi$ and $\widehat{d}_{\omega\kappa m}$ annihilates the corresponding antifermion, while $\psi^{c}$ denotes the charge-conjugated mode function. Our conventions and the explicit anti-commutation relations are summarized in Appendix~\ref{app:quantization-fermion}.

Using the quantized field (\ref{eq:quantization-of-psi}) together with the canonical anti-commutation relations, the fermion bilinear fluctuation is given by
\begin{align}
\label{eq:radialFermion2}
    \langle \overline{\psi}\psi \rangle  = -\frac{1}{2 \pi r^2}  \sum_{\kappa} |\kappa| \int {\rm d}\omega \left(|f_{\omega \kappa}(r)|^2 - |g_{\omega \kappa}(r)|^2\right).
\end{align}

Therefore, the fluctuation is entirely determined by the radial profile functions, which solve the coupled equations in Eq.~(\ref{eq:radialFermion}). 
To proceed, we adopt the same source profile as in the scalar case, Eq.~(\ref{eq:sourcedistribution}). 
Imposing continuity of the wavefunction at $r=R$ leaves us with one overall normalization constant. Similar to the scalar case, this constant must be fixed by the quantization condition. The solution to Eq.~(\ref{eq:radialFermion}) for $r>R$ then takes the form:
\begin{align}
    f_{\omega \kappa} &= \sqrt{\frac{1}{\pi}} \sqrt{\frac{\omega + m_\psi}{k_\text{out}}}\left(k_\text{out}r\right)\left[j_{\ell}(k_{\text{out}} r) \cos \delta_\kappa +  y_{\ell}(k_{\text{out}} r) \sin \delta_\kappa\right],\label{eq:radialFermion-f}  \\
    g_{\omega \kappa} &=  \sgn(\kappa)\sqrt{\frac{1}{\pi}}\sqrt{\frac{k_\text{out}}{\omega+ m_\psi}}\left(k_\text{out}r\right)\left[j_{\ell'}(k_{\text{out}} r) \cos \delta_\kappa +  y_{\ell'}(k_{\text{out}} r) \sin \delta_\kappa)\right],\label{eq:radialFermion-g}
\end{align}
where the phase shift is given by:
\begin{align}
\tan \delta_\kappa = \dfrac{\mathcal{Q}\,j_{\ell'}(k_\text{in}R) j_\ell(k_\text{out}R) - j_{\ell}(k_\text{in}R) j_{\ell'}(k_\text{out}R)}{j_{\ell}(k_\text{in}R) y_{\ell'}(k_\text{out}R) - \mathcal{Q}\,j_{\ell'}(k_\text{in}R) y_{\ell}(k_\text{out}R)}\;, \qquad \mathcal{Q} \equiv \frac{k_\text{in} (\omega + m_\psi)}{k_\text{out}(\omega + m_\psi + m_{\rm M})}\;.\label{eq:phaseshiftgeneralfermion}
\end{align}
Note that $k_{\rm in}$ and $k_{\rm out}$ are given by the same expression in Eq~(\ref{eq:kinkout}) with the replacement $m_\phi$ for $m_\psi$. The derivations of Eqs.~(\ref{eq:radialFermion-f})-(\ref{eq:phaseshiftgeneralfermion}) are provided in Appendix~\ref{app:radial-fermion}. 

It is important to emphasize that, for fermions, the phase shift in Eq.~(\ref{eq:phaseshiftgeneralfermion}) depends explicitly on the bare mass $m_\psi$, whereas in the scalar case the bare mass $m_\phi$ enters only implicitly through $k_{\rm in}$ and $k_{\rm out}$ in Eq.~(\ref{eq:phaseshift}). Since we are interested in the fluctuation outside the source, we will always work in the regime where $m_\psi R \ll 1$ is satisfied.

Plugging Eqs.~(\ref{eq:radialFermion-f})-(\ref{eq:radialFermion-g}) into Eq.~(\ref{eq:radialFermion2}), we arrive at the field fluctuation for $r>R$:
\begin{align}
   \Delta\langle  \overline{\psi}\psi \rangle = -\frac{1}{2\pi^3}\sum_\kappa |\kappa| \int {\rm d\omega}  k_\text{out}\Big[&\left(\omega + m_\psi\right) \big(\left(j_\ell(k_\text{out}r) \cos \delta_\kappa +  y_\ell(k_\text{out}r) \sin \delta_\kappa\right)^2 -j_\ell^2(k_\text{out}r) \big) \nonumber\\
   -&\left(\omega - m_\psi\right)\big(\left(j_{\ell'}(k_\text{out} r) \cos \delta_\kappa +  y_{\ell'}(k_\text{out} r) \sin \delta_\kappa\right)^2 - j_{\ell'}^2(k_\text{out}r) \big)\Big] \;.\label{eq:psibarpsi}
\end{align}
At leading order in $R/r$, only the modes with $\kappa=\pm1$ contribute, corresponding to 
\begin{align}
\kappa=-1:\quad j=\tfrac12,\;\ell=0,\;\ell'=1\;,
\qquad
\kappa=1:\quad j=\tfrac12,\;\ell=1,\;\ell'=0\;. 
\end{align}
Taking $k_{\rm out} R \ll 1$ and $m_\psi R \ll 1$ in Eq.~(\ref{eq:phaseshiftgeneralfermion}), we arrive at the associated phase shifts:
\begin{align}
    \tan \delta_{-1} &= \frac{\epsilon N_\chi}{4 \pi \Lambda^2} k_\text{out} (\omega + m_\psi) F_\psi(m_{\rm M} R) \;, \label{eq:tandelta-1} \\
    \tan \delta_{+1} &= -\frac{\epsilon N_\chi}{4 \pi \Lambda^2}k_\text{out}(\omega- m_\psi) F_\psi(m_{\rm M} R)\;,\label{eq:tandelta+1} 
\end{align}
where we have used the relation $n_\chi = 3N_\chi/(4 \pi R^3)$, and we have defined the fermionic-mediator form factor $F_\psi (z)$ as
\begin{align} \label{eq:FormFactorFermion}
    F_\psi(z)\equiv \frac{3}{z} \left(
        \coth z - \dfrac{1}{z}\right), \qquad z\equiv m_{\rm M}R\;.  
\end{align}
Using $\tan \delta_{\pm 1} \approx \delta_{\pm 1}$, Eq.~(\ref{eq:psibarpsi}) is simplified to
\begin{align}
    \Delta \langle \overline{\psi} \psi \rangle  &= \frac{\delta_{-1}}{\pi^2 r^2}\int_{m_\psi} ^\infty  {\rm d}\omega\frac{\omega +m_\psi}{k_\text{out}} \Bigg[ \left( \frac{1}{2} - \frac{\delta_1}{2\delta_{-1}} \frac{\omega-m_\psi}{\omega + m_\psi} \right)\sin(2 k_\text{out} r ) \nonumber \\
    &+\left( \frac{k_\text{out}}{\omega+ m_\psi} \right)^2 \left( -\frac{1}{(k_\text{out}r)^2 } \sin(2 k_\text{out}r) + \frac{2}{k_\text{out} r} \cos(2 k_\text{out}r) + \sin(2 k_\text{out} r) \right)
    \Bigg]. \label{eq:Deltapsibarpsi}
\end{align}
Substituting the phase shifts (\ref{eq:tandelta-1})-(\ref{eq:tandelta+1}) into the expression above and performing the integrals explicitly (see Appendix~\ref{app:integral-fermion} for details), we obtain:
\begin{align}
\boxed{    
    \Delta \langle \overline{\psi} \psi \rangle = - \frac{3  \epsilon N_\chi m_\psi^2}{4 \pi^3 \Lambda^2} \times \frac{K_2(2 m_\psi r)}{r^3}  \times F_\psi( m_{\rm M} R)\;.\label{eq:psibarpsifinal}
}
\end{align}
The dimensionless parameter $m_{\rm M}R$ appearing in the form factor $F_\psi$ can be rewritten as
\begin{align}
m_{\rm M} R =\epsilon \frac{n_\chi R}{\Lambda^2} = \frac{3\epsilon}{4\pi} \frac{N_\chi}{\Lambda^2 R^2}\;.    
\label{eq:mR-fermion}
\end{align}
Compared to the scalar case in Eq.~(\ref{eq:mR}), the crucial difference is that $m_{\rm M}$ is always real for both $\epsilon = \pm 1$, reflecting the fact that the fermion mass enters linearly in the Dirac equation and therefore never becomes tachyonic.
As in the scalar case, we identify two regimes:
\begin{itemize}
    \item \textbf{Weak-coupling limit}: 
    \[
        \sqrt{N_\chi}/\Lambda \ll R 
        \quad\Rightarrow\quad 
        |m_{\rm M}|R \ll 1\,;
    \]
    \item \textbf{Strong-coupling limit}: 
    \[
        \sqrt{N_\chi}/\Lambda \gg R 
        \quad\Rightarrow\quad 
        |m_{\rm M}|R \gg 1\,.
    \]
\end{itemize}
In the remainder of this section, we analyze these two regimes separately.

\subsection{Weak-coupling limit}
For $|m_{\rm M}|R\ll1$, we expand the form factor in Eq.~(\ref{eq:FormFactorFermion}) using
\begin{align}
  \coth z  = \frac{1}{z} + \frac{z}{3} - \frac{z^3}{45} + {\cal O}(z^5)\;,
\end{align}
which leads to 
\begin{align}
F_\psi (z) = 1 - \frac{z^2}{15}+{\cal O}(z^4)\;.    
\end{align}
Then, substituting Eq.~(\ref{eq:psibarpsifinal}) into Eq.~(\ref{eq:V2psi}) leads to the two-fermion potential:
\begin{equation}
\label{eq:V2psiweak}
    V_{2\psi} (r) = - \frac{3 N_\chi m_\psi^2}{4 \pi^3 \Lambda^4} \frac{K_2(2 m_\psi r)}{r^3}\left(1-\frac{z^2}{15}+{\cal O}(z^4)\right).
\end{equation}

The leading term agrees with the known expression obtained from the Feynman diagram approach in the single-particle limit $N_\chi \to 1$~\cite{Fichet:2017bng}. In the massless limit $m_\psi r  \ll 1$, it exhibits the expected scaling behavior, $V_{2\psi}\sim 1/r^5$.

The next-to-leading term, $-z^2/15$, quantifies the deviation from the Feynman diagram result. Since the sign of the coupling only flips the sign of $z$, and since the correction depends on $z^2$, it is therefore identical for both $\epsilon=\pm 1$. Consequently, the quantum force is suppressed relative to the Born-approximation limit for either choice of the sign.

\subsection{Strong-coupling limit}
\label{subsec:fermionstrong}

We move to the strong-coupling regime. Since the argument in Eq.~(\ref{eq:FormFactorFermion}) is always real and the form factor $F_\psi(z)$ is an even function of $z$, the strong-coupling limit corresponds to taking $z\gg 1$ for both $\epsilon = \pm 1$:
\begin{align}
F_\psi(z\gg 1) = \frac{3}{z} \;.    
\end{align}
Thus, in this regime, the quantum force is suppressed by a factor of $3/(m_{\rm M}R)$ relative to the weak-coupling limit. 

In contrast to the scalar case, fermionic mediators do not exhibit any resonant enhancement at large coupling: the force is purely suppressed. A useful way to see the absence of resonant enhancement for fermionic mediators is to ask whether a zero-energy bound state can form in the interior region. For a bound state, the fermion must be oscillatory inside the source and evanescent outside, which requires
\begin{align}
    k_{\rm in}^2 = \omega^2 - (m_\psi + m_{\rm M})^2 > 0\;, \qquad
    \omega^2 < m_\psi^2\;.
\end{align}
Taking the threshold limit $\omega \to m_\psi^-$, the interior condition becomes
\begin{align}
    -2 m_\psi < m_{\rm M} < 0\;.
\end{align}
So a zero-energy bound state would require an attractive induced mass with $|m_{\rm M}| < 2 m_\psi$. However, in the IR regime relevant for the quantum force, we impose $m_\psi R \ll 1$, which then implies
\begin{align}
    |m_{\rm M}| R < 2 m_\psi R \ll 1\,.
\end{align}
This is incompatible with the strong-coupling condition $|m_{\rm M}| R \gg 1$. We therefore conclude that no zero-energy bound state can exist in the parameter space where the fermionic mediator would otherwise be strongly coupled, and hence, no resonant enhancement arises. The fermion-mediated quantum force is purely screened relative to the weak-coupling result.

\section{Discussions}
\label{sec:pheno}

In this section, we discuss some implications of our results and some possible generalizations. 

\subsection{Other quantum forces}

In the paper, we discussed two examples of quantum forces: one mediated by a scalar, see Eq.~(\ref{eq:LS}), and one mediated by a fermion, see Eq.~(\ref{eq:LF}), both of which are sourced by a single species of fermionic matter.

There are, of course, many natural generalizations. One may consider different mediators (in particular, photons and gravitons), alternative couplings and Dirac structures, or sources composed of multiple fermion species or even scalar constituents.

While most of these extensions require detailed and lengthy calculations to obtain the results, for several classes of generalizations, the results can be extracted straightforwardly from our analysis.

One such case is a source made of real scalars rather than fermions. Let the source field be a real scalar $\Phi$ with mass $m_\Phi$. In this case, the interaction terms in Eqs.~(\ref{eq:LS}) and (\ref{eq:LF}) can be modified accordingly,
\begin{equation}
\frac{\epsilon}{2\Lambda}\, \phi^{2} J \;\to\; \frac{\epsilon \lambda}{2}\, \phi^{2} J'\;, 
\qquad
\frac{\epsilon}{\Lambda^{2}}\, \overline{\psi}\psi\, J \; \to \;
\frac{\epsilon}{\Lambda'}\, \overline{\psi}\psi\, J'\;, \qquad
J' \equiv \Phi^{2}\,.\label{eq:replace}
\end{equation}
The corresponding results can be obtained from Eqs.~(\ref{eq:phisqoutsidefinal}) and (\ref{eq:psibarpsifinal}) by making the simple replacement:
\begin{align}
\frac{1}{\Lambda} \to \frac{\lambda}{m_\Phi}\;,\qquad
\frac{1}{\Lambda^2} \to \frac{1}{m_\Phi\Lambda'}\;.\label{eq:replace2}
\end{align}
Note in particular that in this case the scalar coupling is renormalizable and is controlled by the dimensionless parameter $\lambda$. Requiring perturbativity implies that $\lambda$  must be small.

\subsection{Factorization of the non-perturbative effect}
\label{subsec:factorization}

In all the cases we discussed, the field fluctuation induced by the source factorizes into two pieces: (1) a radial part that reproduces the weak-coupling results obtained using the Feynman diagram approach, and (2) a form factor $F$ that encodes the finite structure and coupling strength of the source (weak or strong). Parametrically,
\begin{align}
\boxed{
\text{Potential = Feynman-diagram Potential $\times$ Form Factor}\;.
}
\end{align}

This factorization arises because we work in the far-field regime $r \gg R$. In this limit, the field is insensitive to the microscopic structure of the source, so the radial dependence reduces to the universal point-particle form, while all information about the density profile and strong or weak coupling dynamics is compressed into a single form factor that depends entirely on $z \equiv m_{\rm M}R$. This reflects a general principle familiar from non-relativistic effective field theories, where physical observables can be factorized into a short-distance (UV, perturbative) part and a long-distance (IR, non-perturbative) part~\cite{Bodwin:1994jh,Lepage:1997cs}.

It is worth emphasizing that, although the explicit form factors we derive depend on the specific source profiles and specific examples studied here, the existence of this factorization is expected to hold in all cases. The qualitative structure does not depend on whether the mediator is scalar or fermionic, nor on the particular Lorentz structure of its couplings, nor on the specific matter profile. As long as we only consider the effect far away from the source, the factorization should hold. 

What does depend on the model is the precise functional form of the form factor $F$. It is sensitive to several ingredients, including the mediator spin (which determines the order of the differential equation), the Lorentz structure of the interaction, and the spatial profile of the source. All of them modify the medium-induced mass and the corresponding boundary-matching conditions. These details determine the quantitative suppression or enhancement encoded in $F$, while the factorization itself remains robust.

\subsection{Physical origin of the suppression factors}
\label{eq:subsec:screening}

\begin{figure}
    \centering
    \includegraphics[width=0.7\linewidth]{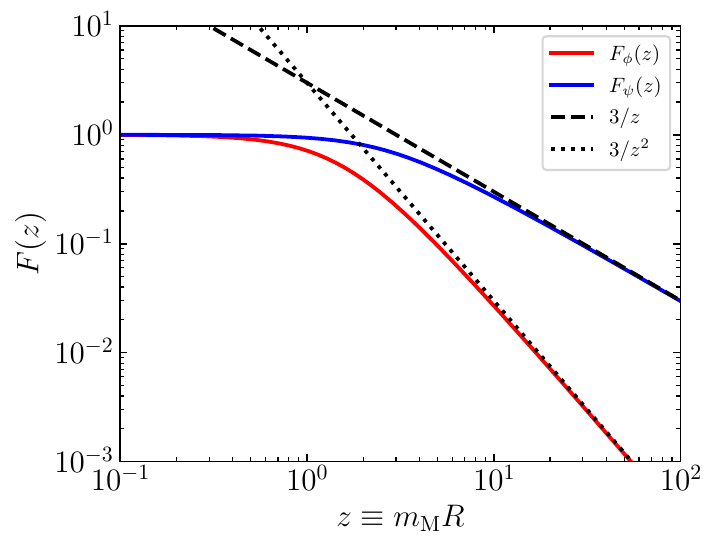}
    \caption{\label{fig:formfactor} Comparison between the scalar-mediator form factor $F_\phi(z)$ [defined in Eq.~(\ref{eq:formfactorscalardef})] and the fermionic-mediator form factor $F_\psi(z)$ [defined in Eq.~(\ref{eq:FormFactorFermion})] in the regime where $z\equiv m_{\rm M}R > 0$. Both form factors approach $1$ in the weak-coupling limit $z \ll 1$, while in the strong-coupling limit $z \gg 1$ they scale as $3/z^{2}$ and $3/z$, respectively. }
\end{figure}

The two concrete form factors derived in this work, $F_\phi(z)$ and $F_\psi(z)$, are compared in Fig.~\ref{fig:formfactor}: both approach unity in the small-$z$ limit, while they are suppressed as $3/z^{2}$ and $3/z$ in the large-$z$ limit, respectively.
Below, we provide a quantitative explanation for the origin of these suppression factors in the strong-coupling limit. In this discussion, we restrict to $\epsilon = + 1$ for the scalar mediator and $\epsilon = \pm 1$ for the fermionic mediator.  

When $m_{\rm M}R \gg 1$, the mediator acquires a large effective mass inside the source, so only a thin shell of thickness $\sim m_{\rm M}^{-1}$ near the surface can contribute to fluctuations that propagate outward.  
This yields a universal geometric suppression factor
\begin{align}
\frac{\text{shell volume}}{\text{bulk volume}} \sim  \frac{4\pi R^{2} m_{\rm M}^{-1}}{(4\pi/3)R^{3}}
= \frac{3}{m_{\rm M}R}\;.    
\end{align}
This geometric suppression factor is common to both scalar and fermionic mediators. What distinguishes the two cases is the order of their EOMs.

Fermionic mediators satisfy a first-order Dirac equation, for which only the spinor itself is required to be continuous at $r=R$.  
Consequently, the geometric-shell effect is the only source of suppression, leading to $F_\psi(z \gg 1) = 3/(m_{\rm M}R)$.

For scalar mediators, the radial Klein-Gordon equation is second order, and both $\phi$ and $\phi'$ must match continuously at $r=R$.  
Inside the source, the heavy-mass solution varies exponentially, so 
$\phi'_{\rm in}(R)/\phi_{\rm in}(R) \sim m_{\rm M}$, 
whereas outside it varies only over the much longer scale $k_{\rm out}^{-1}$.  
The derivative matching, therefore, forces an additional cancellation between the exterior $j_0$ and $y_0$ components, introducing an extra suppression factor $\sim 1/(m_{\rm M}R)$.  
Combining both effects yields $F_\phi(z \gg 1) = 3/(m_{\rm M}R)^{2}$.
Thus, the stronger suppression of scalar-mediated quantum forces reflects the second-order nature of the Klein-Gordon equation and the resulting boundary constraints on the radial wavefunction.

\subsection{Connection between strong-coupling regime and non-perturbativity}
\label{subsec:strongcoupling-nonperturbative}
A key advantage of our approach is that, by avoiding Feynman-diagram expansions, our method can be readily applied to the strong-coupling regime. In the $r\gg R$ limit, which is the regime of interest throughout this work, the potential is factorized into an $r$-dependent part and a source-dependent part that we refer to as the form factor. The $r$-dependent part coincides with the result obtained in the diagrammatic approach, while all deviations from that limit are captured entirely by the $r$-independent form factor.  

This deviation is governed by a single dimensionless parameter, $z$. For the examples considered in this paper, we have
\begin{align}
z \equiv m_{\rm M}R\;,\qquad \text{where}\;
\begin{cases}
z^2 = \dfrac{3\epsilon}{4\pi} \dfrac{N_\chi}{\Lambda R}  & \;\;\text{scalar mediator \& fermionic source}\;, \\[10pt]
z = \dfrac{3\epsilon}{4\pi}\dfrac{N_\chi}{\Lambda^2 R^2} & \;\;\text{fermionic mediator \& fermionic source}\;,\\[10pt]
z^2 = \dfrac{3\epsilon}{4\pi} \dfrac{\lambda N_\Phi}{m_\Phi R}  & \;\;\text{scalar mediator \& scalar source}\;, \\[10pt]
z = \dfrac{3\epsilon}{4\pi}\dfrac{N_\Phi}{\Lambda' m_
\Phi
R^2} & \;\;\text{fermionic mediator \& scalar source}\;,
\end{cases}
\label{eq:z-to-coupling}
\end{align}
where the last two lines correspond to the replacement rules introduced in Eqs.~(\ref{eq:replace})-(\ref{eq:replace2}) when the source particle is a real scalar $\Phi$.

Our result reduces to the standard Feynman diagram result in the  $|z|\ll 1$ limit.
Since $z$ is directly connected to the microscopic mediator-matter coupling, one can then map the strong-coupling regime ($|z|\gg 1$) accessed by our method to the non-perturbative regime of the underlying field theory. 

The simplest way to see this connection is to examine the case represented by the third line of Eq.~(\ref{eq:z-to-coupling}). This corresponds to a renormalizable interaction in which both the mediator and the source particles are scalars.
Perturbativity requires the dimensionless coupling $\lambda \lesssim 4\pi$. 
Using Eq.~(\ref{eq:z-to-coupling}), we have
\begin{align}
|z|^2 = \frac{\lambda}{4\pi} \frac{3 N_\Phi}{m_\Phi R}\;.    
\end{align}
Setting $N_\Phi=1$ and noting that the minimal spatial extent of a single $\Phi$ particle is roughly $R\sim 1/m_\Phi$, we obtain the simple implication
\begin{align}
|z| \gg 1  \quad \implies \quad \lambda \gg 4\pi \cdot {\cal O}(1)\;.
\end{align}
Therefore, the strong-coupling regime corresponds to a non-perturbative coupling between the mediator and source fields in a renormalizable field theory.

There is, however, an alternative interpretation. One may insist that the fundamental theory remains perturbative, $\lambda \ll 4\pi$, but consider a macroscopically large source. In this case, the effective coupling becomes
\begin{align}
\lambda_{\rm eff} \equiv \frac{3N_\Phi}{m_\Phi R}\cdot\lambda\;, 
\end{align}
which can satisfy $\lambda_{\rm eff}\gg 4\pi$ even when $\lambda$ itself is small. This observation provides an effective approach to probing non-perturbative effects in the IR regime:
In the IR limit ($r\gg R$), a non-perturbative theory with an effective coupling $\lambda_{\rm eff} \gg 4\pi$ in vacuum behaves similarly to a perturbative theory with a fundamental coupling $\lambda\ll 4\pi$ in a dense environment. The latter can be treated analytically using our method.

While we have explicitly analyzed only two representative examples, we expect this correspondence to hold in more general theories.
\subsection{The superposition principle}
\label{subsec:superposition}
Quantum forces differ from classical forces in that they intrinsically violate the superposition principle. Fundamentally, this is because quantum forces arise from the nonlinear coupling of the mediator field to the matter field. Classical forces, on the other hand, arise from terms linear in the mediator field.

For classical forces, such as the Coulomb force, the potential generated by many source particles is simply the linear sum of the potentials from each particle:
\begin{align}
V_c (r) = Q\times\frac{e}{4\pi r}\;,    
\end{align}
where $e$ is the electromagnetic coupling, and $Q$ is the total charge in the source. 

At the quantum level, this linearity is no longer exact. Increasing the charge $Q$ effectively increases the characteristic energy scale $\mu$, causing the coupling $e$ to run according to the renormalization-group (RG) equation: $\beta(e)\equiv \mu \frac{\partial e}{\partial\mu}$. If we parametrize this running in terms of the total charge by defining $\beta_Q(e) \equiv Q \frac{\partial e}{\partial Q}$, then the Coulomb potential satisfies
\begin{align}
Q\frac{\partial V_c}{\partial Q} = V_c\left(1+ \frac{\beta_Q(e)}{e}\right).\label{eq:RGVc}
\end{align}
Since $e(\mu)$ depends only logarithmically on the scale, the violation of superposition is likewise only logarithmic. In the IR limit, the running becomes negligible, and the classical force effectively satisfies the superposition principle.

On the other hand, quantum forces violate the superposition principle in a fundamentally different way. As shown explicitly in Secs.~\ref{sec:scalar} and \ref{sec:fermion}, and discussed more generally in Sec.~\ref{subsec:factorization}, the quantum force in the IR regime ($r\gg R$) takes a factorizable form
\begin{align}
V_q(r) = Q\times V_1(r) \times F\left[z(Q,R,\lambda)\right],    
\end{align}
where $V_1$ is the potential generated by a single source particle, computable using standard Feynman diagrams, and $Q$ denotes the total effective ``charge'' of the source. 

The key nonlinearity arises from the form factor $F$, which depends on the dimensionless parameter $z\equiv m_{\rm M}R$, itself a function of the effective charge $Q$, the source size $R$, and the fundamental coupling strength $\lambda$ between the mediator and matter fields. (For the examples in Eq.~(\ref{eq:z-to-coupling}), $Q$ is roughly proportional to the number of particles inside the source, so $z \sim \sqrt{Q}$ for scalar mediators and $z\sim Q$ for fermionic mediators.)
Because the form factor has a nontrivial dependence on $z$, the resulting quantum force does not scale linearly with $Q$, even at large distances where classical forces obey superposition.

Similar to Eq.~(\ref{eq:RGVc}), one can derive an RG-like equation for the quantum force that quantifies the departure from superposition:
\begin{align}
Q\frac{\partial V_q}{\partial Q} &=  Q V_{1} F + Q^2 \frac{\partial V_1}{\partial \lambda}\frac{\partial \lambda}{\partial Q}F +  Q^2 V_{1}\frac{\partial F}{\partial Q}\nonumber\\
& = V_q\left(1 + \beta_\lambda\frac{1}{V_1}\frac{\partial V_1}{\partial \lambda}  + \beta_z\frac{F'}{F} \right),\label{eq:RGVq}
\end{align}
where we have defined
\begin{align}
\beta_\lambda \equiv Q \frac{\partial\lambda}{\partial Q}\;,\qquad
\beta_z \equiv Q \frac{\partial z}{\partial Q}\;, \qquad F'\equiv \frac{{\rm d}F(z)}{{\rm d}z}\;.
\end{align}

In Eq.~(\ref{eq:RGVq}), the second term captures a violation of superposition analogous to the classical force case in Eq.~(\ref{eq:RGVc}): it arises from the RG running of the microscopic coupling $\lambda$ and therefore produces only logarithmic corrections.
The third term, however, is qualitatively different and unique to quantum forces. It originates from the nontrivial dependence of the form factor $F(z)$ on the effective parameter $z(Q,R,\lambda)$, which encodes the collective effects of the source on the mediator field and leads to a parametrically stronger violation of the superposition principle than the mild logarithmic running present in classical forces.

Only in the leading small-$z$ limit does $F\to 1$, thereby restoring linearity and recovering the superposition principle. In general, however, the nontrivial dependence of $F$ on $z$ (and therefore on $Q$) implies that quantum forces violate the superposition principle in a fundamentally non-logarithmic way.

\subsection{Comparison with the previous literature}

Our framework treats vacuum fluctuations as an effective background, providing a unified method for computing quantum forces both in vacuum and in the presence of a background of on-shell mediator particles. As has been extensively explored in the literature, quantum forces have an interesting feature that they can be significantly modified by a background of mediator particles/fields~\cite{Horowitz:1993kw,Ferrer:1998ju,Ferrer:1998rw,Ferrer:1999ad,Hees:2018fpg,Fukuda:2021drn,Ghosh:2022nzo,Banerjee:2022sqg,Blas:2022ovz,Day:2023mkb,VanTilburg:2024xib,Barbosa:2024pkl,Ghosh:2024qai,Bauer:2024yow,Banerjee:2025dlo,delCastillo:2025rbr,Grossman:2025cov,Cheng:2025fak,Gan:2025nlu,Ferrante:2025lbs,Ittisamai:2025oxf,Gan:2025icr}.
Intuitively, this effect can be understood as a type of ``wake force''~\cite{VanTilburg:2024xib}. Another perspective is that one of the particles in the loop is effectively replaced by an on-shell particle from the background~\cite{Ghosh:2022nzo}.
Several methods exist to calculate quantum forces modified by a background. One approach involves thermal field theory~\cite{Horowitz:1993kw,Ferrer:1998ju,Ferrer:1998rw,Ferrer:1999ad,Ghosh:2022nzo,Blas:2022ovz,Barbosa:2024pkl,Ghosh:2024qai,Cheng:2025fak,Ferrante:2025lbs,Ittisamai:2025oxf}. Another is the picture of coherent scattering between the test particle and the background particle~\cite{VanTilburg:2024xib, Grossman:2025cov}.  A third approach uses the expectation value of the squared background field~\cite{Hees:2018fpg,Banerjee:2022sqg,Bauer:2024yow,Banerjee:2025dlo,delCastillo:2025rbr,Gan:2025nlu,Gan:2025icr}. This last method is closer in spirit to the classical field picture, as the expectation value acts as the source generating the force. Throughout this paper, we do not assume the presence of a background of mediator particles. Therefore, although we start from similar quadratic couplings, there are no relevant background-induced (wake) forces in our case.

Our method for obtaining the field solution of the scalar mediator shares some similarity with the method used in \cite{Hees:2018fpg}, but differs in two crucial aspects. First, we do not assume the existence of a classical $\phi$ background, while in~\cite{Hees:2018fpg}, $\phi$ itself is the dark matter particle. This leads to different boundary conditions for the EOM of the mediator field. Second, the observable studied in \cite{Hees:2018fpg} is intrinsically classical, proportional to the square of the background field value, $\sim \phi^2$. In our case, this effect vanishes because there is no classical source. The relevant observable arises purely at the quantum level due to the fluctuation of the mediator field, $\sim \langle \phi^2 \rangle$. Consequently, the resulting force exhibits a different scaling behavior from that found in a classical background.

Screening phenomena analogous to those found in our strong-coupling regime have appeared in other contexts. Screening of scalar fields in dense environments is well known in modified-gravity scenarios, most notably in chameleon models and their variants~\cite{Khoury:2003aq,Khoury:2003rn,Hinterbichler:2010es,Hinterbichler:2011ca,Brax:2018grq}. 
A similar suppression effect has also been noted in \cite{Hees:2018fpg,Banerjee:2022sqg,Day:2023mkb,Gan:2025nlu,Gan:2025icr} under the assumption that $\phi$ constitutes a classical dark matter background.

Our results show that an analogous type of screening arises even in the absence of any classical background, and it occurs for both scalar and fermionic mediators.
When the mediator-source coupling is strongly repulsive, the field fluctuation generated inside the source becomes effectively trapped, preventing it from propagating outward.
Consequently, a test particle located outside the source experiences a suppressed quantum force. 
In the IR regime $r\gg R$, this suppression factor becomes independent of the distance.

\subsection{Deviation from the non-relativistic approximation}
\label{subsec:relativistic}

In this work, we have assumed that both the test particle and the source particles inside the $\chi$ cluster are non-relativistic, as is typically done in the standard Feynman diagram approach within the Born approximation. However, there exist physically relevant situations in which this non-relativistic assumption breaks down. For example, electrons in heavy atoms move with relativistic velocities ($v \sim Z\alpha$), and a $\chi$ cluster may itself be a bound state composed of relativistic constituents --- much like nucleons, which contain relativistic quarks.

In contrast to the traditional method, our approach can be extended to the relativistic regime in a straightforward manner, for both the test particle and the source. If the test particle $\chi$ is relativistic, Eq.~\eqref{eq:EOMchi} continues to hold, though it is no longer appropriate to interpret $\epsilon\langle\phi^{2}\rangle/\Lambda$ as an effective ``potential'' in the usual non-relativistic sense.

The method can also be generalized to sources containing relativistic $\chi$ particles. In this case the approximation $\langle\overline{\chi}\chi\rangle\approx n_\chi$ is no longer valid. Instead, one must use the fully relativistic expression (see, e.g., Appendix A of~\cite{Smirnov:2022sfo}):
\begin{align}
    \langle \overline{\chi}\chi \rangle\equiv \int\frac{{\rm d}^{3}\vecp}{(2\pi)^{3}}\frac{m_{\chi}}{E_{\vecp}}f_{\chi}(\vecp)\;,\label{eq:chibarchirelativistic}
\end{align}
where $E_\vecp \equiv \sqrt{m_\chi^2 + \vecp^2}$ and $f_{\chi}(\vecp)$ is the phase-space distribution of $\chi$. 

Neutron stars provide a natural example: neutrons are bound by gravity and become increasingly relativistic as the star approaches the Oppenheimer-Volkoff limit. In such bound systems, even though the constituents are relativistic, the overall $\chi$ cluster is static, and therefore sources a static profile of $\langle\phi^{2}\rangle$. This may have phenomenological implications for axion physics if one identifies $\phi$ with the axion field and $\chi$ with the neutron field, since a similar quadratic coupling $\phi^2\overline{\chi}\chi$ arises in chiral effective theory~\cite{Hook:2017psm,Fukuda:2021drn}. A systematic investigation of the relativistic regime is left for future work.

\subsection{Possible astrophysical applications}
\renewcommand\arraystretch{1.3}
\begin{table}[t]
\centering
\begin{tabular}{|c|c|c|c|}
\hline
Source & $n_N~[\text{fm}^{-3}]$ & $R~[\text{km}]$ & $z \equiv G_F n_N R$ \\
\hline
Earth        & $3.3\times 10^{-15}$ & $6.4\times 10^{3}$ & $10$ \\
\hline
Sun          & $8.4\times 10^{-16}$ & $7.0\times 10^{5}$ & $2.7\times 10^{2}$ \\
\hline
White dwarf  & $6.0\times 10^{-10}$ & $7.0\times 10^{3}$ & $1.9\times 10^{6}$ \\
\hline
Neutron star & $1.6\times 10^{-1}$ & $1.0\times 10^{1}$ & $7.3\times 10^{11}$ \\
\hline
\end{tabular}
\caption{Representative nucleon number densities $n_N$, radius $R$, and the dimensionless parameter $z \equiv G_F n_N R$ for various astrophysical objects.}
\label{tab:zparameter}
\end{table}
\renewcommand\arraystretch{1.0}

Finally, let us estimate how typical astrophysical environments influence the quantum force. As a concrete example, consider the case in which the matter field $\chi$ is the nucleon and the fermionic mediator $\psi$ is the neutrino. The dimensionless parameter that characterizes the degree of non-perturbativity is
\begin{align}
z \equiv m_{\rm M}R \sim  G_F n_N R\;,
\end{align}
where $n_N$ is the nucleon number density. The numerical values of $z$ in typical astrophysical objects are summarized in Table~\ref{tab:zparameter}. This indicates that the neutrino force among typical astrophysical objects significantly violates the superposition principle.

The number density of nucleons in a neutron star is extremely high, around the nuclear saturation density, $n_N\approx 0.16 /{\rm fm}^3$. Taking the radius $R\sim 10{\rm km}$, one obtains $z\sim G_F n_N R\sim 10^{12} \gg 1$, implying that neutrino forces in neutron stars should be in the strong-coupling regime. As previously discussed (see Sec.~\ref{subsec:fermionstrong}), such a high value of $z$ causes a strong screening effect on neutrino forces. A naive summation of the $r^{-5}$ potential in Eq.~\eqref{eq:nu-force} sourced by each nucleon is invalid, since the dense nuclear matter not only sources neutrino forces, but also suppresses the spread of the forces. One could ask, with the nuclear saturation density, to what scale individual contributions from nucleons can be summed directly.  This corresponds to the screening length scale $R_{\rm screen}\sim 1/m_{\rm M}\sim 10 {\rm nm}$, which is still much larger than the mean distance between nucleons, $R_{\rm mean}\sim n_N^{-1/3}\sim 1.8 {\rm fm}$. Hence, approximately $(R_{\rm screen}/R_{\rm mean})^3\sim 10^{20}$ nucleons can be summed up as if they are in vacuum. So the screening effect can be neglected within a single heavy nucleus with ${\cal O}(100)$ nucleons, but not within a neutron star. The stability of a dense star under the potentially large many-body neutrino force has been a subject of debate in the literature~\cite{Fischbach:1996qf,Smirnov:1996vj,Abada:1996nx,Kiers:1997ty}. While we did not include the Dirac structure of the weak interaction, our results suggest that the strong screening effect of the neutrino force may play an important role in mitigating these effects and could help maintain the stability of the star.

\section{Conclusions}
\label{sec:conclusion}

In this paper, we introduce a new method to calculate the potential generated by quantum forces. The basic idea is to consider a cluster of source particles and to determine the long-distance potential by computing the expectation value of the square of the mediator field. For classical forces one evaluates the expectation value of the field itself, but for quantum forces the relevant quantity is its square.

Traditionally, one-loop Feynman diagrams have been used to calculate quantum forces, an approach that assumes weak couplings in order to be valid. Our method allows us to compute the force also in the strong-coupling regime. In particular, we reproduce the Feynman-diagram results in the leading weak-coupling limit. Moreover, we calculate the small corrections in the weak-coupling regime, and extend the result to the case of strong couplings between the mediator and source particles.

Our main finding is that, in the strong-coupling regime, the superposition principle no longer holds in the simple way familiar from the electric force. One cannot decompose a source into its building blocks and add the contributions of each part independently. Mathematically, this deviation appears in a factorized form: the potential equals the perturbative result (which assumes superposition) multiplied by a function that parametrizes the deviation. We refer to this function as the form factor.

This deviation can either suppress or enhance the potential, depending on the sign of the interaction between the source particles and the mediators. For repulsive interactions, the form factor leads to a suppression, whereas for attractive interactions, it produces an enhancement. Based on this argument, we expect a suppression of the Casimir-Polder force, which is mediated by two photons, and an enhancement of the graviton-mediated quantum force. However, confirming these expectations requires a full calculation.

\renewcommand\arraystretch{1.3}
\begin{table}[ht]
    \centering
    \begin{tabular}{|c | c | c |}
        \hline
         & Scalar Mediator & Fermionic Mediator \\
        \hline
        Full Expression & (\ref{eq:phisqoutsidefinal}) & (\ref{eq:psibarpsifinal}) \\
        \hline
        Weak-Coupling Limit & (\ref{eq:V2phiweak}) & (\ref{eq:V2psiweak}) \\
        \hline
        Form Factor & (\ref{eq:formfactorscalardef}) & (\ref{eq:FormFactorFermion}) \\
        \hline
        Screening & $3/z^2$ & $3/z$ \\
        \hline
        Resonance & (\ref{eq:resonanceRatio}) & - \\
        \hline
    \end{tabular}
    \caption{Summary of our main results on quantum forces mediated by scalars and fermions.}
    \label{tab:summary}
\end{table}

\renewcommand\arraystretch{1.0}

Our main quantitative results are summarized in Table \ref{tab:summary}.
There are several directions in which these results can be extended, especially to other types of mediators. In particular, it would be valuable to determine how large an enhancement is possible. A very significant enhancement could provide a way to probe forces that are currently purely theoretical.

\section*{Acknowledgement}
We thank Matthias Neubert and Maxim Perelstein for helpful discussions. YG and BY are supported by the NSF grant PHY-2309456. XJX is supported in part by the National Natural
Science Foundation of China under grant No.~12141501 and also by
the CAS Project for Young Scientists in Basic Research (YSBR-099).

\begin{appendix}
\section{Quantization of the mediator field}
In this appendix, we present the canonical quantization of both scalar and fermionic fields in the spherical wave basis. For each case, the (anti)commutation relations of the creation/annihilation operators, as well as those of the fields themselves, are explicitly specified. Given these sets of (anti)commutation relations, the completeness relations for the radial mode functions follow uniquely. These completeness relations play a crucial role: they are required to fully determine the field solutions and, consequently, the magnitude of the resulting quantum fluctuations.
\label{app:quantization}
\subsection{Scalar mediator}
\label{app:quantization:scalar}
In the spherical basis, we quantize the scalar field as:
\begin{equation}
    \widehat{\phi}(\vecr,t) =  \int {\rm d}\omega \sum_{\ell,m} \frac{u_{\omega \ell}(r)}{r}\left(\widehat{a}_{\omega \ell m}^{} Y_{\ell m}(\theta, \varphi)\,e^{- i \omega t} + \widehat{a}^\dagger_{\omega \ell m}Y^*_{\ell m}(\theta, \varphi)\,e^{i \omega t} \right).\label{eq:quantization-of-phi-app}
\end{equation}
The creation and annihilation operators satisfy the following commutator relations:
\begin{align}
    \left[ \widehat{a}_{\omega\ell m} , \widehat{a}^\dagger_{\omega' \ell' m'}\right] &=  \frac{1}{\omega}\delta(\omega- \omega') \delta_{\ell \ell'} \delta_{m m'}\;, \nonumber\\
    \left[\widehat{a}_{\omega\ell m} , \widehat{a}_{\omega' \ell' m'}\right] &=\left[\widehat{a}^\dagger_{\omega\ell m} , \widehat{a}^\dagger_{\omega' \ell' m'}\right] = 0\;.
\end{align}
The field $\widehat{\phi}$ also satisfies the following equal-time canonical commutation relation:
\begin{align}
    [\widehat{\phi}(\mathbf{r}), \dot{\widehat{\phi}}(\mathbf{r'})] = i \delta^3\left(\mathbf{r}- \mathbf{r'}\right),
\end{align}
which follows from the following completeness relations:
\begin{align}
    \int {\rm d}\omega\,u_{\omega \ell}(r) u_{\omega \ell}(r') & = \frac{1}{2}\delta(r- r')\;,\\
    \label{comRelationScalar} 
    \sum_{\ell,m} Y_{\ell m}(\theta, \varphi) Y_{\ell m}(\theta',\varphi') &= \delta(\cos \theta - \cos \theta') \delta(\varphi -\varphi')\;.
\end{align}
The dual radial relation of Eq.~(\ref{comRelationScalar}),
\begin{align}
    &\int {\rm d}r\,u_{\omega \ell}(r) u_{\omega' \ell}(r)  = \frac{1}{2}\delta(\omega - \omega')\;.
\end{align}
serves as the normalization condition that uniquely determines the overall factor of the radial solutions $u_{\omega \ell}$ for any $\omega$ and $\ell$.

\subsection{Fermionic mediator}\label{app:quantization-fermion}
As discussed in the main text, the fermionic field is quantized using spherical spinor harmonics:
\begin{align}
    \widehat{\psi}(\mathbf{r}, t) = \sum_{\kappa, m}\int {\rm d} \omega \left( \widehat{b}_{\omega\kappa m}\psi_{\omega \kappa m} (\mathbf{r})\,e^{-i \omega t}  + \widehat{d}^\dagger_{\omega\kappa m} \psi^c_{\omega \kappa m}(\mathbf{r})\,e^{i \omega t}  \right),
\end{align}
with the spatial wavefunctions given by:
\begin{align}
    \psi_{\omega\kappa m} (\mathbf{r}) = \frac{i}{r} \begin{pmatrix}
    f_{\omega \kappa}(r)\,\Omega_{\kappa m}(\hat{\mathbf r}) \\[4pt]
    i\,g_{\omega \kappa}(r)\,\Omega_{-\kappa m}(\hat{\mathbf r})
  \end{pmatrix}, \quad \psi^c_{\omega\kappa m}
  = C\,\overline{\psi_{\omega\kappa m}}^{\,T}\;,
  \end{align}
where $C = i \gamma^2 \gamma^0 $ is the charge-conjugation matrix in the Dirac representation, $\Omega$ are spinor spherical harmonics defined in Eq.~(\ref{eq:spinorsphericalharmoinicsdef}), and $f$ and 
$g$ are radial profile functions. The non-vanishing anti-commutation relations are
\begin{align}
\{\widehat{b}_{\omega\kappa m},\,\widehat{b}^\dagger_{\omega'\ell'm'}\} =
  \{\widehat{d}_{\omega\kappa m},\,\widehat{d}^\dagger_{\omega'\ell'm'}\}
   &= \delta(\omega-\omega')\,\delta_{\ell \ell'}\,\delta_{mm'}\;.
\end{align}
The field satisfies the equal-time canonical anti-commutation relation:
\begin{equation}
  \{\widehat\psi(\mathbf r),\widehat\psi^\dagger(\mathbf r')\}
  = \delta^3(\mathbf r-\mathbf r')\,\mathbf 1_{4\times4}\;.
\end{equation}
This follows from the completeness relation for all energy-modes:
\begin{align} \label{eq:comRelationFermion}
    \int {\rm d}\omega \left[f_{\omega \kappa} (r)  f^*_{\omega \kappa} (r') + g_{\omega \kappa} (r)  g^*_{\omega \kappa} (r')\right]  & =\delta (r- r')\;.
\end{align} together with the completeness of the spinor spherical harmonics:
\begin{align}
  \sum_{\kappa, m}\Omega_{\kappa m}(\hat{\mathbf r})\Omega_{\kappa m}^\dagger(\hat{\mathbf r}')
   &= \delta(\widetilde{\Omega}-\widetilde{\Omega}')\,\mathbf 1_{2\times2}\;,
\end{align}
where $\widetilde{\Omega}$ on the right-hand side denotes the solid angle.
The dual radial relation of Eq.~(\ref{eq:comRelationFermion}), given by
\begin{align}\label{eq:comDualFermion}
    \int {\rm d}r\left[f_{\omega \kappa} (r)  f^*_{\omega' \kappa} (r) + g_{\omega \kappa} (r)  g^*_{\omega' \kappa} (r) \right]  & =\delta (\omega- \omega')\;,
\end{align}
provides the normalization needed to uniquely fix the radial functions $f_{\omega \kappa}$ and $g_{\omega \kappa}$  for any $\omega$ and $\kappa$.
  
\section{Calculation of the radial functions}
 In this appendix, we show how to solve the radial equations for both scalar and fermionic fields in the presence of $\chi$ clusters. For each case, we begin by writing down the general solutions in terms of spherical Bessel functions of the first and second kinds. We then use the spherical-cow setup laid out in the main text to obtain two piecewise solutions, one in each region (inside and outside the cow). We then match the solutions at the boundary by imposing continuity (and differentiability, for the scalar case). The matching conditions allow us to determine all but one overall normalization constant. By fixing this remaining constant using the completeness relations given in the last appendix, we obtain a fully determined radial functions for both the scalar and fermionic fields.
\subsection{Scalar mediator}
\label{app:radial-scalar}
The radial equation of motion for a real scalar field $\phi$ with mass $m_{\phi}$ is given by:
\begin{equation}
\left[ \partial_r^2 - \frac{\ell(\ell + 1)}{r^2} + \omega^2- m_{\phi}^2\right] u_{\omega \ell}(r) = 0\;. \label{eq:radialScalar-app}
\end{equation}
The general solution to this equation is given by the spherical Bessel functions of the first and second kinds:
\begin{align}
    u_{\omega \ell}(r) = a_1 j_\ell({kr)} + a_2 y_\ell(kr)\;,
\end{align}
where $a_1$ and $a_2$ are some constants and $k^2 \equiv \omega^2 - m_{\phi}^2$. 

For a spherical cow of radius $R$, the effective mass squared of the field is $m^2_\phi + m^2_{\rm M}$ for $r \leq R$, and $m^2_\phi$ for $r>R$. Defining  $k_\text{in}^2 = \omega^2 - (m_\phi^2 + m_{\rm M}^2) $ and $k_\text{out}^2 = \omega^2 -m_\phi^2$, the solution to Eq.~(\ref{eq:radialScalar-app}) for a spherical cow can be constructed from the solutions in the region $r\leq R$ and the region $r>R$:
    \begin{align}
u_{\omega \ell}(r) &=
\begin{cases}
A_{\text{in}}   j_{\ell}(k_{\text{in}} r) , &\text{for} \quad  r \leq R\;, \\
A_{\text{out}}  j_{\ell}(k_\text{out} r) + B_{\text{out}}  y_{\ell}(k_{\text{out}} r) &\text{for} \quad r > R\;,
\end{cases}
\end{align}
where $A_\text{in}$, $A_{\rm out}$, $B_{\rm out}$ are unknowns to be determined.
Imposing continuity and differentiability at $r=R$, we find that:
\begin{align}
 A_{\text{out}} &= A_{\text{in}} (k_{\text{out}} R)^2 \left[
j_\ell(k_{\text{in}} R) y_\ell'(k_{\text{out}} R)
- \frac{k_{\text{in}}}{k_{\text{out}}} j_\ell'(k_{\text{in}} R) y_\ell(k_{\text{out}} R)
\right]\;, \\
B_{\text{out}} &= A_{\text{in}} (k_{\text{out}} R)^2 \left[
\frac{k_{\text{in}}}{k_{\text{out}}} j_\ell'(k_{\text{in}} R) j_\ell(k_{\text{out}} R)
- j_\ell(k_{\text{in}} R) j_\ell'(k_{\text{out}} R)
\right] .
\end{align}
To obtain $A_{\rm in}$, we need to make use of the completeness relation. Consider the asymptotic behavior of $u_{\omega \ell}(r)$:
\begin{align}
    \lim_{r\to \infty} u_{\omega \ell} (r) = \frac{A_{\text{out}}}{k_{\text{out}}} \sin \left(k_{\text{out}}r - \frac{\pi \ell}{2}\right) - \frac{B_{\text{out}}}{k_{\text{out}}} \cos \left(k_{\text{out}}r - \frac{\pi \ell}{2}\right).
\end{align}
Let $A_{\rm out} = N k_{\rm out} \cos \delta_\ell$ and $B_{\rm out} = N k_{\rm out} \sin \delta_\ell$, in a compact form, we can write:
\begin{align}
    \lim_{r\to \infty} u_{\omega \ell} (r) = N \sin \left(k_\text{out}r - \frac{\pi \ell}{2} - \delta_\ell\right).
\end{align}
Now, using the completeness relation, 
\begin{align}
    \int {\rm d}r\,u_{\omega \ell}(r)u_{\omega' \ell}(r) = \frac{1}{2}
\delta(\omega- \omega')\;,
\end{align} we obtain
\begin{equation}
    N = \sqrt{\frac{1}{\pi}\frac{\omega}{k_\text{out}}}\;.
\end{equation}

\subsection{Fermionic mediator}
\label{app:radial-fermion}
The coupled radial equations for a Dirac fermion with mass $m_\psi$ is given by:
\begin{align}
    \frac{\rm d}{{\rm d}r} \begin{pmatrix}
       f_{\omega \kappa} (r) \\  g_{\omega \kappa}(r)
    \end{pmatrix} = 
    \begin{pmatrix}
        -\kappa/r & \omega+m_\psi \\
        -(\omega-m_\psi) & \kappa/r
    \end{pmatrix}
    \begin{pmatrix}
         f_{\omega \kappa} (r) \\  g_{\omega \kappa}(r)
    \end{pmatrix},
\end{align}
The general solution to this coupled differential equation is given by the spherical Bessel function of the first and the second kind:
\begin{align}
    f_{\kappa \omega }(r) &=  r \left( a_1 j_\ell(kr) + a_2 y_{\ell} (kr)  \right), \\
    g_{\kappa \omega }(r) &= \sgn(\kappa) \frac{k r}{\omega+ m_\psi} \left( a_1 j_{\ell'} (kr) + a_2 y_{\ell'} (kr)  \right),
\end{align}
where $a_1$ and $a_2$ are some constants, and $k^2  \equiv \omega^2 - m_\psi^2$. 

For a spherical cow, the effective mass is $m_\psi + m_{\rm M}$ for $r<R$, and $m_\psi$ for $r>R$. The general solution is given by the two following piecewise functions:
\begin{align}
f_{\omega \kappa}(r) &=
\begin{cases}
A_{\text{in}}  \, r j_{\ell}(k_{\text{in}} r)  &\text{for}\quad r \leq R\;, \\
A_{\text{out}} \, r j_{\ell}(k_\text{out} r) + B_{\text{out}} \, r y_{\ell}(k_{\text{out}} r) &\text{for}\quad r > R\;,
\end{cases}\\
g_{\omega \kappa}(r) &= \sgn(\kappa)r
\begin{cases}
\dfrac{k_\text{in} }{\omega + m_\psi + m_{\rm M}}A_{\text{in}} \,  j_{\ell'}(k_{\text{in}} r)  &\text{for}\quad  r \leq R \;, \\
\dfrac{k_\text{out} }{\omega + m_\psi}\left(A_{\text{out}}   j_{\ell'}(k_{\text{out}} r) + B_{\text{out}}y_{\ell'}(k_{\text{out}} r)\right) &\text{for}\quad r > R\;,
\end{cases}
\end{align}
where $A_\text{in}$, $A_{\rm out}$, $B_{\rm out}$ are constants to be determined, $k_\text{in}^2 = \omega^2 - (m_\psi+m_{\rm M})^2$, and $k_\text{out}^2 = \omega^2 -m_\psi^2$. Matching the solutions at the boundary $r=R$, we can obtain $A_\text{out}$ and $B_\text{out}$:
\begin{align}
A_{\text{out}}
&=
\frac{A_{\text{in}}}{\Delta}
\left[
      j_{\ell}(k_{\text{in}} R)\, y_{\ell'}(k_{\text{out}} R)
      \;-\;
      \frac{k_{\text{in}}(\omega+m_\psi)}{k_{\text{out}}(\omega+m_\psi+m_{\rm M})}\,
      j_{\ell'}(k_{\text{in}} R)\, y_{\ell}(k_{\text{out}} R)
\right], \nonumber
\\[6pt]
B_{\text{out}}
&=
\frac{A_{\text{in}}}{\Delta}
\left[
      \frac{k_{\text{in}}(\omega+m_\psi)}{k_{\text{out}}(\omega+m_\psi+m_{\rm M})}\,
      j_{\ell}(k_{\text{in}} R)\, j_{\ell'}(k_{\text{out}} R)
      \;-\;
      j_{\ell'}(k_{\text{in}} R)\, j_{\ell}(k_{\text{out}} R)
\right],
\end{align}
where 
\begin{equation}
\Delta \;\equiv \;
  j_{\ell}(k_{\text{out}} R)\,
  y_{\ell'}(k_{\text{out}} R)
  \;-\;
  y_{\ell}(k_{\text{out}} R)\,
  j_{\ell'}(k_{\text{out}} R)\;.
\end{equation}
Again, we define $A_{\rm out} = N k_{\rm out} \cos \delta_\ell$ and $B_{\rm out} = N k_{\rm out} \sin \delta_\ell$. Like the scalar case, we can use the completeness relation, Eq.~(\ref{eq:comDualFermion}), to obtain
\begin{align}
   N = \sqrt{\frac{\omega +  m_\psi} {\pi k_\text{out}}}\;.
\end{align}
\section{Calculation of integrals relevant for the field fluctuation}
In this appendix, we provide additional details on the evaluation of integrals relevant for computing the field fluctuations induced by a spherical $\chi$ cluster for both scalar and fermionic mediators.
For the scalar case, we show how an apparently divergent integral can be evaluated exactly using contour methods, yielding a standard Bessel function expression. For the fermionic case, we present the explicit results for the integrals entering the field fluctuation and, crucially, show how they combine into a compact form involving known special functions.

\subsection{Scalar mediator}
\label{app:contour}

\begin{figure}
    \centering
    \includegraphics[width=0.7\linewidth]{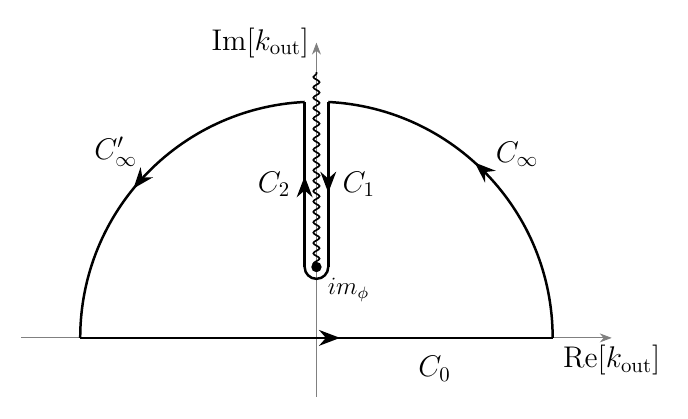}    \caption{The contour of integration in the complex $k_{\rm out}$-plane. The path along the real axis is denoted by $C_0$. The contour is deformed to wrap around the branch cut starting at $i m_\phi$, with vertical segments $C_1$ and $C_2$.}
    \label{fig:contourPath}
\end{figure}

In the main text, we showed that the scalar field fluctuation outside the source can be reduced to a compact expression involving a single integral (see Eq.~(\ref{eq:phisqoutside})):
\begin{align}
    I &\equiv \int_0^\infty \frac{k_\text{out}}{\sqrt{k_\text{out}^2+m_\phi^2}} \sin\left(2 k_\text{out} r\right){\rm d}k_\text{out}\;.\label{eq:Idef}
\end{align}
Mathematically, this integral does not converge as $k_{\rm out}\to \infty$ on the real axis. However, from the physical point of view, UV modes do not contribute to the field fluctuation or to the resulting quantum force. In the following, we show how to remove this apparent divergence using a contour-integration method, which renders the expression finite without introducing any regulator.

Since the integrand is even in $k_\text{out}$, we can extend the range of integration to $(-\infty, \infty)$ while also replacing the sine function with an exponential:
\begin{align}
     I &=\frac{1}{2}\,\Im \left[\int_{-\infty}^\infty f(k_{\rm out})\, {\rm d}k_\text{out}\right], \quad \text{where} \;\; f(k_\text{out}) \equiv \frac{ k_\text{out}}{\sqrt{k_\text{out}^2+m_\phi^2}}\;e^{ 2ik_\text{out}r}\;.
     \label{eq:Iexpform}
\end{align}
To evaluate the integral, we analytically continue $k_{\rm out}$ into the complex plane and perform the integral along the path in Fig~\ref{fig:contourPath}. The integrand has branch points at $i m_\phi$ and a branch cut along the imaginary axis extending from $im_\phi$ to $i \infty$. We avoid this branch cut by taking the path along $C_1$ and $C_2$, while closing the contour at infinity with $C_\infty$ and $C'_{\infty}$. 

The closed contour $\mathcal{C}= C_0 + C_1 + C_2 + C_\infty+ C_{-\infty}$ encloses no poles. Therefore, by Cauchy's theorem,
\begin{align}
    \int_\mathcal{C} f(k_{\rm out}) \;{\rm d}k_{\rm out} = 0\;.
\end{align}
The contributions from $C_{\infty}$ and $C_{-\infty}$ vanish; therefore, the integral along the real axis can be determined by the contribution along the branch cut:
\begin{align}
    \int_{C_0} f(k_{\rm out}) \;{\rm d}k_{\rm out}=  - \left( \int_{C_1} f(k_{\rm out}) \, {\rm d} k_{\rm out} + \int_{C_2} f(k_{\rm out}) \, {\rm d}k_{\rm out} \right). \label{eq:C0integral}
\end{align}
Along the branch cut, we parameterize $k_{\rm out} = i t$. 
On the right side of the cut ($C_1$), $k_{\rm out} = it + \varepsilon$ (with $\varepsilon \to 0^+$), so $\sqrt{k_{\rm out}^2+m_\phi^2} = i\sqrt{t^2-m_\phi^2}$.
On the left side of the cut ($C_2$), $k_{\rm out} = it - \varepsilon$, and due to the phase wrap around the branch point, the square root changes sign: $\sqrt{k_{\rm out}^2+m_\phi^2} = -i\sqrt{t^2-m_\phi^2}$.

Substituting these into Eq.~(\ref{eq:C0integral}):
\begin{align}
    \int_{-\infty}^\infty \frac{ k_\text{out}\,e^{ 2ik_\text{out}r}}{\sqrt{k_\text{out}^2+m_\phi^2}} {\rm d}k_\text{out} 
    &= - \left[ \int_{\infty}^{m_\phi} \frac{(it) e^{-2tr}}{i\sqrt{t^2-m_\phi^2}} (i\,{\rm d}t) + \int_{m_\phi}^{\infty} \frac{(it) e^{-2tr}}{-i\sqrt{t^2-m_\phi^2}} (i\,{\rm d}t) \right]\nonumber \\
    &= 2i \int_{m_\phi}^\infty \frac{t e^{-2tr}}{\sqrt{t^2-m_\phi^2}} {\rm d}t\;.
\end{align}
This integral is well defined and convergent, and it can be rewritten in terms of a modified Bessel function of the second kind:
\begin{equation}
 2i\int_{m_\phi}^\infty \frac{t e^{-2tr}}{\sqrt{t^2-m_\phi^2}} {\rm d}t = 2i m_\phi K_1(2m_\phi r)\;.
\end{equation}
Substituting this back into the expression for $I$ in Eq.~(\ref{eq:Iexpform}), we obtain:
\begin{align}
    I &= \frac{1}{2}\,\Im\left[ 2i m_\phi K_1(2m_\phi r) \right] =   m_\phi K_1(2m_\phi r)\;.
\end{align}

\subsection{Fermionic mediator}
\label{app:integral-fermion}
We now present the explicit forms of the integrals entering the field fluctuation in Eq.~(\ref{eq:Deltapsibarpsi}) after substituting the phase shifts in Eqs.~(\ref{eq:tandelta-1})-(\ref{eq:tandelta+1}):
\begin{align}
    \Delta \langle \overline{\psi} \psi \rangle = \frac{\epsilon N_\chi}{4 \pi^3 \Lambda^2 r^2} \left( I_1 + m_\psi^2 I_2 -\frac{1}{r^2}I_2 +\frac{2}{r}I_3 + I_4 \right) F_\psi( m_{\rm M} R)\;, \label{eq:Deltapsibarpsi-app}
\end{align}
where the relevant integrals read:
\begin{align}
    I_1 &\equiv \int_0 ^\infty  k \sqrt{k^2 + m_\psi^2} \sin(2 kr)\,{\rm d}k = - \frac{m_\psi^2}{2r} K_2(2 m_\psi r)\;, \label{eq:I1}\\
    I_2 &\equiv \int_0 ^\infty \frac{k}{\sqrt{k^2 + m_\psi^2}} \sin(2 kr)\,{\rm d}k = m_\psi K_1( 2m_\psi r)\;,\\
    I_3 &\equiv\int_0 ^\infty \frac{k^2}{\sqrt{k^2 + m_\psi^2}} \cos(2kr)\,{\rm d}k = \frac{1}{2} m_\psi^2 G_{1,3}^{2,1}\left(m_\psi^2 r^2 \Big|
    \begin{array}{c}
     -\frac{1}{2} \\
     -1,0,\frac{1}{2} \\
    \end{array}
    \right), \\
    I_4 &\equiv \int_0^\infty \frac{k^3}{\sqrt{k^2 + m_\psi^2}} \sin (2 k r)\,{\rm d}k = \frac{1}{2} m_\psi^3 G_{1,3}^{2,1}\left(m_\psi^2 r^2 \Big|
    \begin{array}{c}
     -1 \\
     -\frac{3}{2},\frac{1}{2},0 \\
    \end{array}
    \right),\label{eq:I4}
    \end{align}
where $G$ is the Meijer-G function defined by \cite{Prudnikov1990v3}:
\begin{align*}
G^{m,n}_{p,q}\!\left( x \,\bigg|\, \begin{array}{l}
a_1,\ldots,a_p \\
b_1,\ldots,b_q
\end{array} \right)
\equiv
\frac{1}{2\pi i}
\int_{\gamma_L}
\frac{
\displaystyle \prod_{j=1}^{m} \Gamma(b_j + s)\;
        \prod_{j=1}^{n} \Gamma(1 - a_j - s)
}{
\displaystyle \prod_{j=n+1}^{p} \Gamma(a_j + s)\;
        \prod_{j=m+1}^{q} \Gamma(1 - b_j - s)
}
\, x^{-s}\, {\rm d}s\;,
\end{align*}
with $\Gamma(s)$ denoting the gamma function. Note that all the integrals $I_1$-$I_4$ can be computed using the same contour-integration method as the integral in Eq.~(\ref{eq:Idef}).

We further notice that the integrals in Eqs.~(\ref{eq:I1})-(\ref{eq:I4}) satisfy the following identity:
\begin{align}
    m_\psi^2 I_2 - \frac{1}{r^2}I_2 + \frac{2}{r}I_3 + I_4 = 5I_1 \;.
\end{align}
Substituting it back into Eq.~(\ref{eq:Deltapsibarpsi-app}) and using Eq.~(\ref{eq:I1}) we obtain
\begin{align}
    \Delta \langle \overline{\psi} \psi \rangle &= \frac{3 \epsilon N_\chi I_1}{2 \pi^3 \Lambda^2 r^2} F_\psi( m_{\rm M} R) =  - \frac{3  \epsilon N_\chi m_\psi^2}{4 \pi^3 \Lambda^2} \times \frac{K_2(2 m_\psi r)}{r^3}  \times F_\psi( m_{\rm M} R)\;.
\end{align}

\section{Calculation of higher-$\ell$ modes}
\label{app:higherell}
In this appendix, we calculate the correction to the quantum force from $\ell > 0$ modes for the scalar case. According to Eq.~(\ref{eq:differencescalar}), the field fluctuation outside the source is given by
\begin{align}
    \Delta \langle \phi^2 \rangle &= \frac{1}{\pi}
\sum_{\ell} \dfrac{2\ell + 1}{4\pi} \int_{0}^\infty \frac{{\rm d} k_\text{out} k_\text{out}^2}{\sqrt{k^2_\text{out}+ m^2_\phi}} \; \nonumber\\
& \times\left[\frac{\tan^2 \delta_\ell}{1+ \tan^2 \delta_\ell}\left( 
- j_\ell^2 ( k_\text{out}r) + y_\ell^2(k_\text{out} r)\right) + \frac{2\tan \delta_\ell}{1+ \tan^2 \delta_\ell}j_\ell( k_\text{out} r) y_\ell( k_\text{out} r) \right],
\end{align}
with $\tan\delta_\ell$ given by Eq.~(\ref{eq:phaseshift}).
For small $\delta_\ell$, we have
\begin{align}
    \Delta \langle \phi^2 \rangle &\approx \frac{2}{\pi}
\sum_{\ell} \dfrac{2\ell + 1}{4\pi} \int_{0}^\infty \frac{ {\rm d} k_\text{out} k_\text{out}^2}{\sqrt{k^2_\text{out}+ m^2_\phi}} \delta_\ell\,j_\ell( k_\text{out} r) y_\ell( k_\text{out} r) \;,
\end{align}
with
\begin{align}
\delta_\ell \approx \frac{\pi}{2^{2\ell+1}\Gamma\left(\ell+\frac{1}{2}\right) \Gamma\left(\ell+\frac{3}{2}\right)}\frac{I_{\ell+\frac{3}{2}}\left(m_{\rm M}R\right)}{I_{\ell-\frac{1}{2}}\left(m_{\rm M}R\right)} \left(k_{\rm out}R\right)^{2\ell + 1}\;,    
\end{align}
where $\Gamma$ is the Euler Gamma function and $I_\ell$ is the modified Bessel function of the first kind.

For small $\delta_\ell$, the above expression is simplified as
\begin{align}
\Delta\langle \phi^2 \rangle = \frac{1}{2\pi^2 r^2} \sum_\ell  C_\ell(z) \left(\frac{R}{r}\right)^{2\ell+1} {\cal I}_\ell(m_\phi r)\;,    
\end{align}
where the coefficient is a function of $z\equiv m_{\rm M}R$:
\begin{align}
C_\ell(z) &\equiv \frac{\left(2\ell+1\right)\pi}{2^{2\ell+1}\Gamma\left(\ell+\frac{1}{2}\right) \Gamma\left(\ell+\frac{3}{2}\right)}\frac{I_{\ell+\frac{3}{2}}\left(z\right)}{I_{\ell-\frac{1}{2}}\left(z\right)}\;,
\end{align}
and the integral is defined as 
\begin{align}
{\cal I}_\ell(m_\phi r) &\equiv \int_0^\infty {\rm d}x \frac{x^3}{\sqrt{x^2+(m_\phi r)^2}}\,x^{2\ell} j_\ell(x) y_\ell(x)\;.   
\end{align}

\subsection{$\ell = 1$ mode}
In what follows, we explicitly compute the field fluctuation for the $\ell=1 $ mode:
\begin{align}
    \Delta \langle \phi^2 \rangle_{\ell = 1} = \frac{3}{4 \pi^2} \int_{0}^\infty \frac{k_\text{out}^2 {\rm d} k_\text{out}}{\sqrt{k^2_\text{out}+ m^2_\phi}} \left(\left( j_1(k_\text{out} r) \cos \delta_1 + y_1 (k_\text{out} r) \sin \delta_1 \right)^2 - j_1^2(k_\text{out} r) \right)\;
\end{align}
Note that for $\ell=1$, the spherical Bessel functions can be written as
\begin{align}
    j_1(x) = \frac{\sin x}{x^2} - \frac{\cos x}{x}\;, \quad y_1(x) = - \frac{\cos x}{x^2} - \frac{\sin x}{x}\;.
\end{align}
Therefore, we can write the field fluctuation as
\begin{align}
    \Delta \langle \phi^2 \rangle_{\ell = 1} = -\frac{3}{4 \pi^2}& \int_0^\infty \frac{k_\text{out}^2 {\rm d} k_\text{out}}{\sqrt{k^2_\text{out}+ m^2_\phi}}\nonumber \\
     & \times \left[\left(\frac{1}{(k_\text{out} r)^4} - \frac{1}{(k_\text{out}r)^2} \right)\left(\sin (2 k_\text{out} r) \frac{\tan \delta_1}{ 1+ \tan^2\delta_1} - \cos(2 k_\text{out} r) \frac{\tan^2 \delta_1}{ 1+ \tan^2\delta_1}\right)\right.  \nonumber\\
    &- \left. \frac{2}{(k_\text{out} r)^3}
    \left(\cos (2 k_\text{out} r) \frac{\tan \delta_1}{ 1+ \tan^2\delta_1} + \sin(2 k_\text{out} r) \frac{\tan^2 \delta_1}{ 1+ \tan^2\delta_1}\right)
     \right].
\end{align}
Now, the phase shift is given by
\begin{align}
    \tan \delta_1 = \frac{z^2}{45}\left(k_\text{out}  R \right)^3 F_{\phi,1}(m_{\rm M} R)\;,
\end{align}
where the $p$-wave form factor is found to be
\begin{align}
 F_{\phi, 1}(z) = \frac{15}{z^4} \left(3+z^2- 3z \coth z \right). 
\end{align}
For definiteness, we take $\epsilon=+1$ so that $z$ is real. The form factor $F_{\phi,1}$ has the limit:
\begin{align}
F_{\phi,1}(z\ll 1) &= 1 -\frac{2 z^2}{21} +{\cal O}(z)^4 \;,\\
F_{\phi,1}(z\gg 1) &= \frac{15}{z^2}\;.
\end{align}
Using $\tan \delta_1 \approx \delta_1$, the field fluctuation is reduced to:
\begin{align}
\Delta \langle \phi^2 \rangle_{\ell = 1} 
&= \frac{- z^2}{60 \pi^2} 
   \int_0^\infty 
   \frac{\delta_1\, k_{\text{out}}^{\,2} {\rm d} k_{\rm out}}
        {\sqrt{k_{\text{out}}^{\,2} + m_\phi^2}}
   \left[
      \left(
        \frac{1}{(k_{\text{out}} r)^4}
        - \frac{1}{(k_{\text{out}} r)^2}
      \right)
      \sin(2 k_{\text{out}} r)
      \;-\;
      \frac{2}{(k_{\text{out}} r)^3}
      \cos(2 k_{\text{out}} r)
   \right]
\nonumber\\
&= 
\frac{N_\chi}{160 \pi^3} \frac{m_\phi}{\Lambda}\frac{R^2}{r^4}\, 
F_{\phi,1}(m_{\rm M} R)
\;\times\;
\Bigg[
m_\phi^2 r^2\,
G_{1,3}^{2,1}\!\left(
   m_\phi^2 r^2 \,\Bigg|\,
   \begin{array}{c}
      -1 \\
      -\frac{3}{2},\,\frac12,\,0
   \end{array}
\right)
\nonumber\\[4pt]
&\hspace{3.4cm}
+~
2 m_\phi r\,
G_{1,3}^{2,1}\!\left(
   m_\phi^2 r^2 \,\Bigg|\,
   \begin{array}{c}
      -\tfrac12 \\
      -1,\,0,\,\tfrac12
   \end{array}
\right)
\;-\;
2 K_1(2 m_\phi r)
\Bigg].
\end{align}
In the massless limit, $m_\phi r \ll 1$, we obtain a $1/r^5$ scaling behavior,
\begin{align}
    \Delta \langle \phi^2 \rangle_{\ell = 1} = -\frac{N_\chi }{64 \pi^3 \Lambda r^3} \frac{R^2}{r^2} F_{\phi,1}(m_{\rm M} R)\;.
\end{align}
Therefore, compared to the $\ell=0$ mode, the field fluctuation from the $\ell =1$ mode is suppressed by a factor of $R^2/r^2$ and therefore can be ignored in the far-field limit $r\gg R$.
\end{appendix}

\bibliographystyle{JHEP}
\bibliography{ref}
\end{document}